\documentstyle[12pt]{article}

\makeatletter

\@addtoreset{equation}{section}
\def\section{\@startsection {section}{1}{\z@}{-3.5ex plus -1ex minus
 -.2ex}{2.3ex plus .2ex}{\large\bf\centering}}
\def\subsection{\@startsection{subsection}{2}{\z@}{-3.25ex plus%
 -1ex minus -.2ex}{1.5ex plus .2ex}{\sc}}

\advance \voffset by -1.0in
\advance \hoffset by -0.9in
\def\tr{\mbox{tr}}
\def\cd{\!\cdot\!}
\def\cross{\!\times\!}
\def\bea{\begin{eqnarray}}
\def\eea{\end{eqnarray}}
\def\bpi{{\mbox{\boldmath $\pi$}}}

\def\btau{{\mbox{\boldmath $\tau$}}}

\def\bal{{\mbox{\boldmath $\alpha$}}}
\def\bbet{{\mbox{\boldmath $\beta$}}}
\def\bom{{\mbox{\boldmath $\omega$}}}
\def\bOm{{\mbox{\boldmath $\Omega$}}}
\def\bgam{{\mbox{\boldmath $\gamma$}}}
\def\bxi{{\mbox{\boldmath $\xi$}}}
\def\bet{{\mbox{\boldmath $\eta$}}}

\def\bone{{\mbox{\bf 1}}_2}

\def\bP{{\mbox{\boldmath $P$}}}
\def\bp{{\mbox{\boldmath $p$}}}
\def\bd{{\mbox{\boldmath $d$}}}
\def\be{{\mbox{\boldmath $e$}}}
\def\br{{\mbox{\boldmath $r$}}}
\def\bs{{\mbox{\boldmath $s$}}}

\def\bq{{\mbox{\boldmath $q$}}}
\def\bn{{\mbox{\boldmath $n$}}}
\def\bm{{\mbox{\boldmath $m$}}}
\def\bx{{\mbox{\boldmath $x$}}}
\def\bX{{\mbox{\boldmath $X$}}}
\def\bY{{\mbox{\boldmath $Y$}}}
\def\bu{{\mbox{\boldmath $u$}}}
\def\bv{{\mbox{\boldmath $v$}}}
\def\bR{{\mbox{\boldmath $R$}}}
\def\bS{{\mbox{\boldmath $S$}}}

\def\bI{{\mbox{\boldmath $I$}}}
\def\bJ{{\mbox{\boldmath $J$}}}

\def\bM{{\mbox{\boldmath $M$}}}

\def\bL{{\mbox{\boldmath $L$}}}
\def\L{L^{(1)}}
\def\R{R^{(2)}}
\def\Lbar{\bar{ L}^{(1)}}
\def\Rbar{\bar{ R}^{(2)}}
\def\l{l^{(1)}}
\def\r{r^{(2)}}
\def\O{{\cal O}}
\def\d{\dagger}

\begin{document}
\baselineskip 18pt
\parskip 7pt
\begin{flushright}
DTP 93-29

hep-ph/9308236
\end{flushright}

\begin{center}

\vspace{4cm}
{\huge Dynamics of Moving and Spinning Skyrmions}

\vspace{1.5cm}

{\Large
B.J.Schroers$^{\ast}$\\
Department of Mathematical Sciences\\
South Road\\
Durham DH1 3LE\\
United Kingdom\\}

\vspace{1cm}

\centerline{ published in Z.Phys.C61(1994) 479}

\vspace{1cm}

{\bf Abstract}

\end{center}

{\small
We study  the intermediate- and  long-range forces between moving and spinning
Skyrmions,
 employing
two different methods. One uses a relativised product ansatz
 for the Skyrme fields,
the other models Skyrmions as triplets of scalar dipoles.
The methods lead to the same finite-dimensional
Lagrangian dynamical system which
may be interpreted as a point-particle approximation to Skyrmion
dynamics. We discuss in detail the dynamics in the
so-called attractive channel and the interaction between well-separated and
rapidly spinning Skyrmions,  and point out the resemblance
between the latter and the one-pion exchange potential in nuclear physics.
}

\vfill
$\ast$ e-mail: b.j.schroers@durham.ac.uk
\pagebreak

\section{Introduction}

The Skyrme model is a non-linear, non-integrable  field theory,  defined on 3+1
dimensional \linebreak
Minkowski space, which  has soliton solutions \cite{Skyrme}.
Here we will use the term ``soliton"  to
mean what is also sometimes called a topological soliton: a localised,
finite-energy and smooth solution of
the field equations which is stable for topological reasons.
Examples include lumps in the ${\bf CP}^n$ model, vortices
in the abelian Higgs model and BPS monopoles in Yang-Mills-Higgs theory.
In all cases the solitons have an associated integer winding number which
is a topological invariant and which is also called the soliton number.
Skyrme solitons of soliton number one are referred to as  Skyrmions.
Suitably quantised they are models for physical baryons.

Given the long-standing
difficulties in finding a satisfactory theoretical model for the interaction
of two baryons (the nuclear two-body problem) it is not surprising  that
much effort has been devoted to the study of  the classical and  quantised
interaction of two Skyrmions.
In the absence of integrability one cannot hope to find exact time-dependent
solutions which represent two-Skyrmion scattering or bound states, and one
therefore has
to resort either to numerical  methods or to  some kind of approximation.

One scheme which has been used to great effect  in the study
   of soliton dynamics at low energy is the adiabatic or
moduli space approximation, explained systematically in ref. \cite{M1}.
In this scheme one views the field theory as an infinite-dimensional
Lagrangian dynamical system $({\cal C},L)$, consisting of a configuration space
$\cal C$, the space of all fields at a given time, and a Lagrangian
of the form $L=T-V$, where $T$ is the  kinetic energy functional and
$V$ is the potential energy functional.
One then truncates the theory to a finite-dimensional Lagrangian system
by choosing a finite-dimensional submanifold ${\cal M} \in {\cal C}$ and
restricting $L$ to $\cal M$. The space $\cal M$ is called the moduli space
and should contain all configurations which, on general physical grounds,
are deemed to be relevant at low energy.
In theories with topological solitons the procedure is somewhat simplified
because $\cal C$ decomposes naturally into disjoint sectors ${\cal C}_n$
which are each labelled by the winding number $n$ of the fields
in it. Correspondingly, one may consider each winding number
separately and choose, for given $n$, a truncation ${\cal M}_n$ of
${\cal C}_n$. The elements of ${\cal M}_n$ are called $n$-solitons.
The basic assumption of the moduli space approximation is that there
exist finite-dimensional manifolds ${\cal M}_n$ such that the
solutions of the truncated Lagrangian systems $({\cal M}_n, L_{\mid {\cal
M}_n})$
are, in some sense, good approximations to solutions of the full field
theory.

How does one find  candidates for ${\cal M}_n$?
${\cal M}_1$ is usually defined as the set of all  fields with winding
number one which minimise $V$. Typically, this set includes internal
as well as position parameters.
For $n\geq 2$, ${\cal M}_n$ should include configurations consisting of
$n$ well-separated single solitons with arbitrary internal parameters and
positions, and one therefore requires that
dim ${\cal M}_n =\> n \cross $ dim ${\cal M}_1$.
In certain theories, which are said to be of Bogomol'nyi type,
this  requirement is automatically satisfied if one defines ${\cal M}_n$, like
${\cal M}_1$, as the set of fields in ${\cal C}_n$ which minimise $V$.
It follows that $V$ is constant on ${\cal M}_n$, which means physically that
there are no static forces between solitons of Bogomol'nyi type.
As a result the
truncated dynamics is entirely determined by the induced kinetic
energy. This is generally of the form $T = {1\over 2} g_{ij}\dot q_i
\dot q_j$, where $\{q_i\}$ are coordinates on ${\cal M}_n$ and the matrix
$g_{ij}$, which depends on $q_i$, defines a Riemannian metric on
the moduli space. Thus in the moduli space approximation to the
slow motion of solitons of Bogomol'nyi type the interactive dynamics
of $n$ solitons is modelled by geodesic motion on the moduli space.
This is an adiabatic approximation because it approximates the time evolution
of $n$-solitons by a sequence of static solutions of the field equations.
So far the most interesting and successful application of moduli space
approximation has been to two-soliton dynamics in theories  of
Bogomol'nyi  type.

Skyrme's theory is not of Bogomol'nyi type, but the
static forces between Skyrme solitons are quite weak.
It has therefore been suggested \cite{M1} that Skyrme solitons, like
Bogomol'nyi type solitons, might be amenable to the moduli space
approximation. This requires that one finds suitable moduli spaces for Skyrme's
theory.
${\cal M}_1$, defined as above, is six-dimensional and contains three
translational
and three rotational degrees of freedom. Thus ${\cal M}_2$ should be
12-dimensional.
The set of minima of the potential energy $V$  amongst two-Skyrmions
is an eight-dimensional manifold of
coincident Skyrmions \cite{axi} and
does not include well-separated Skyrmions.
On the other hand, there is a 12-dimensional family of two-Skyrmions,
namely the fields of the product form, first considered by Skyrme in ref.
\cite{Skyrprod}, which
includes
well-separated Skyrmions but does not correctly describe the low-energy
coincident configurations.
In ref. \cite{M1} N.S. Manton suggested a method of constructing a
12-dimensional
manifold ${\cal M}^{12}$ which would include both  the coincident and the
well-separated
Skyrme fields just described.
More recently Atiyah and Manton have shown  that one can obtain good
approximations
to Skyrme fields  of low energy by calculating the holonomy of $SU(2)$
instantons. They obtain a 12-dimensional manifold $\tilde {\cal M}^{12}$ of
two-Skyrmions
in this way and describe its topological and differentiable structure in
ref. \cite{AM}.
They argue further that $\tilde {\cal M}^{12}$ should be a good approximation
to
${\cal M}^{12}$. The next step in the moduli space approximation, which has not
yet been carried out, is to calculate the induced Lagrangian on ${\cal M}^{12}$
or on $\tilde {\cal M}^{12}$.
There are partial results for the potential, but little is known about  the
kinetic energy.

In  the present paper we study the  kinetic energy of
two well-separated Skyrmions which, statically, are well described
by the product ansatz. Physically, this means studying the velocity-dependent
forces between
moving and spinning Skyrmions.
According to the moduli space approximation  one should, in order
to calculate these forces, allow the 12 parameters in the product ansatz
to vary with time and  express the kinetic energy in terms of the 12 parameters
and their derivatives with respect to time. A few of the terms generated in
this way were calculated in ref. \cite{NR} and a more
complete numerical study was undertaken recently \cite{SWA}.
In this paper we will argue that this approach does not lead to the correct
effective Lagrangian.
Rather, one must allow the individual Skyrmions to Lorentz-contract when they
move and modify the product ansatz to take into account this and other
relativistic effects.

To back up our results we re-derive them using a different method.
Our second calculation is based on the observation that, from afar, a
Skyrmion looks like a triplet of point-like scalar dipoles. It has been  known
for
a long time that the potential between two static Skyrmions can be
understood in this way. Using retarded potentials we calculate the Lagrangian
for the interaction
of two spinning and moving triplets of scalar dipoles including all terms
which are at most quadratic in linear and angular velocities.
Similar retarded potential calculations,
which may be interpreted as point-particle approximations to
soliton dynamics, have been carried out for BPS monopoles
\cite{M2} and maximally charged black holes \cite{Ru} and have confirmed
the moduli space approximation for the long-range dynamics.
In our case the Lagrangian for the dipole interaction agrees with the
Skyrme Lagrangian evaluated on the relativised product ansatz.

The aim of this paper is two-fold. Specifically we want to
understand the importance and nature of kinematic effects in
two-Skyrmion dynamics. More generally we address  the wider question of how to
approximate classical  soliton interactions. In this introduction we have
described the moduli space approximation  in some detail because it
provides a coherent and general attempt to answer  that question.
At the end of this paper  we will
give a critical assessment of the moduli space
approximation in the light of our results.

\section{The Skyrme Model}

The fundamental field of Skyrme's  theory is a map
\bea
U: \mbox{ M}^4 \mapsto SU(2)
\eea
where M$^4$ is Minkowski space with signature
$(+,-,-,-)$.
We  denote elements of M$^4$ by $x$, with coordinate $x^{\mu}$, $\mu =0,1,2,3$.
Sometimes we also write $x=(t,\bx)$.
It is often useful to parametrise $U$ as
\bea
U(x) =  {1\over f_{\pi} } \left(
 \sigma (x) + i{\bpi}(x)\cd{\btau} \right)
\eea
with the constraint $\sigma^2 + \bpi^2 = f_{\pi}^2$.
$\tau^a$, $a=1,2,3$,
are the Pauli matrices, $\pi^a$ the pion fields and $f_{\pi}$ is
the pion decay constant.
The Lagrangian density is best written in terms  of
the pull-back of the left-invariant        form  $U^{\dagger}dU$
\bea
L_{\mu} =  U^{\dagger}\partial_{\mu}U,
\eea
where $\partial_{\mu} = \partial /\partial x^{\mu}$,
or in terms of the right-invariant analogue
\bea
R_{\mu} = U\partial_{\mu}U^{\dagger}.
\eea
It reads
\bea
\label{Lag}
{\cal L} &=& -{f_{\pi}^2\over 4 }\tr(L_{\mu}L^{\mu}) + {1 \over 32 e^2}
\tr(\lbrack L_{\mu},L_{\nu}\rbrack\lbrack L^{\mu},L^{\nu}\rbrack )
\nonumber \\
     &=&  -{f_{\pi}^2\over 4 }\tr(R_{\mu}R^{\mu}) + {1 \over 32 e^2}
\tr(\lbrack R_{\mu},R_{\nu}\rbrack\lbrack R^{\mu},R^{\nu}\rbrack ).
\eea
The constant $e$ is a free parameter of
 the Skyrme model and it is customary to treat $f_{\pi}$ as a free parameter as
well. We will adjust $e$ and $f_{\pi}$ later to fit experimental
data. We fix our units by setting Planck's constant and the speed of light
to 1.

Note the large symmetry group of the action corresponding to (\ref{Lag}).
 It is invariant
under
\bea
 \mbox{Poincar\'e group } \cross SO(4)_{\mbox{\tiny chiral}}
\eea
where
 \bea
SO(4)_{\mbox{\tiny chiral}} \cong {SU(2) \cross SU(2) \over {\bf Z}_2}
\eea
acts on $U(x)$    via left and right multiplication with constant
$SU(2)$ matrices.
The Euler-Lagrange equations take the form of conservation laws:
the currents
\bea
\bar{L}_{\mu} =  L_{\mu} +{1\over 4 e^2 f^2_{\pi}}
\lbrack L_{\nu},\lbrack L^{\nu},L_{\mu}\rbrack \rbrack
\eea
and
\bea
\bar{R}_{\mu} =  R_{\mu} +{1\over 4 e^2 f^2_{\pi}}
\lbrack R_{\nu},\lbrack R^{\nu},R_{\mu}\rbrack \rbrack
\eea
satisfy
\bea
\label{EL}
\partial_{\mu} \bar{L}^{\mu} =\partial_{\mu} \bar{R}^{\mu}  = 0.
\eea

So far we have assumed that the pions are massless.
Physical pions have a small mass of about $m_{\pi} \approx 140$MeV
(we ignore the mass difference between $\pi^0$ and $\pi^{\pm}$).
This can be incorporated into the Skyrme model by adding
\bea
\label{mass}
{1\over 4} f^2_{\pi} m^2_{\pi} \tr(U + U^{\dagger} -2)
\eea
to the Lagrangian.
This term explicitly breaks the chiral symmetry to  \linebreak
$SO(3)_{\mbox{\tiny isospin}}$,  which acts via conjugation with a
constant $SU(2)$ element $A$
\bea
\label{iso}
 U(x) \mapsto  A U(x)A^{\dagger},
\eea
or alternatively, in terms of the associated orthogonal matrix
\bea
\label{orth}
 {\cal A}^{ab} = {1\over 2} \tr (\tau^a A \tau^bA^{\dagger})
\eea
and the pion field,
\bea
\label{rot}
\pi^a(x) \rightarrow {\cal A}^{ab} \pi^b(x).
\eea
In the presence of  the pion mass term the equations of motion become
\bea
\label{ELm}
\partial_{\mu} \bar{L}^{\mu} = -\partial_{\mu} \bar{R}^{\mu}  =
{1\over 2}m^2_{\pi}(U-U^{\dagger}).
\eea
It turns out  that the predictions of the Skyrme model for
the static properties of a single nucleon do not  depend much on the pion
mass, as long as it is small \cite{AN}. For the interaction of two Skyrme
solitons,  however,
the pion mass has  important
physical consequences.
Here we  will  carry out calculations mainly for the case
 $m_{\pi} = 0$ where the theory is more symmetric and the mathematical
manipulations somewhat simpler.  In certain applications we can then
take into account the physical pion mass by general physical arguments.

Viewed
as an infinite dimensional Lagrangian system
the Skyrme model's configuration
space ${\cal C}$ is the
space of
maps
\bea
U : {\bf R}^3 \mapsto    SU(2).
\eea
The Lagrangian
$L = \int d^3x \> {\cal L}$ has the usual form
\bea
\label{SKY}
L = T - V
\eea
where $T$ is the kinetic energy
\bea
\label{KE}
T &=& -{f_{\pi}^2\over 4 }\tr(L_0 L_0) - {1 \over 16 e^2}
       \tr(\lbrack L_0,L_i\rbrack \lbrack L_0,L_i\rbrack )
\eea
  and
\bea
\label{Skyrpot}
V &=& -{f_{\pi}^2\over 4}\tr(L_i L_i) - {1 \over 32 e^2}
       \tr(\lbrack L_j,L_i\rbrack \lbrack L_j,L_i\rbrack )
\eea
is the potential energy functional.
To make $V$ well defined, one needs to impose a finite energy requirement
on elements of ${\cal C}$. In practice one demands that the field $U$
tends to a constant matrix,
which one usually takes to be the identity,  at spatial infinity.
 This allows one to compactify
space to $S^3$ and regard $U$ as an element of $\Pi_3(S^3)\cong {\bf Z }$
with an associated integer degree or winding number.
This number is absolutely conserved during time evolution and  is
identified with the baryon number $B$.
A subset of $\cal C$ whose elements have a definite winding
number $B$ is  denoted by  ${\cal C}_B$.

 The imposition of the asymptotic
condition also has  important consequences for
the symmetry of the theory. The chiral symmetry is broken to
$SO(3)_{\mbox{\tiny isospin}}$.
The
euclidean group $E_3 = SO(3)_{\mbox{\tiny space}}\times {\bf R}^3$, whose
elements $({\cal G},\bX)$ act on fields $U\in {\cal C}$
via pullback:
\bea
U(\bx) \mapsto U({\cal  G}^{-1}(\bx -\bX)),
\eea
and the isopspin rotations
$SO(3)_{\mbox{\tiny isospin}}$ are symmetries of the Lagrangian and
map each sector ${\cal C}_B$ of the configuration space
into  itself.
There is also a discrete symmetry, the combination of parity operations
in space and isospace. Separately these parity operations reverse the sign
of the winding number $B$, but the map
\bea
\label{inversion}
\Pi : U(\bx) \mapsto U^{\d}(-\bx)
\eea
leaves $B$ unchanged.
Thus  for each baryon number $B$ the symmetry group of the Lagrangian system
$({\cal C}_B,L_{|{\cal C}_B})$ is
\bea
\label{symmetry}
S= \Pi \cross E_3 \cross SO(3)_{\mbox{\tiny isospin}}.
\eea

\section{Dynamics of a Single Skyrmion}

We briefly review the standard approximations used in discussing single
Skyrmion dynamics and assess their validity.

The usual way to obtain a particular solution of the static
field equations in the sector $B=1$ is to make the hedgehog ansatz
\bea
\label{hedge}
U_{H} (\bx ) = \exp(i f(r) \hat{ \bx} \cd \btau) \quad  r= |\bx|, \quad
\> \hat{ \bx} = {\bx \over r}
\eea
with the boundary
conditions  $f(0) = \pi$ and $f(\infty ) = 0$
to ensure that $U_{H}$ has   winding number 1.
Minimising the
potential energy
$V$ restricted to Skyrme fields of hedgehog type leads
to a second order  ordinary differential     for $f$, given in ref. \cite{ANW},
which can easily be solved numerically.
In  the following $f$ will always stand for that solution and
the corresponding field $U_{H}$ will be referred to as the
standard Skyrmion.
We are particularly  interested in the behaviour of $f$ for large $r$,
which is
\bea
\label{asymf}
 f \sim {\lambda \over r^2} + O({1\over r^8})
\eea
where
$\lambda$ is numerically found to be $\> 2.16 / e^2 f_{\pi}^2\>$.
There is strong numerical evidence that the standard Skyrmion
 is also  a minimum of $V$ among all fields in the  $B=1$ sector
and that all minima in  this sector can be obtained by acting
on the standard Skyrmion with the symmetry group $S$.
This group is 9-dimensional, but
when it  acts on a field of
the hedgehog form  spatial and isospin rotations are equivalent.
Hence
the orbit of the standard Skyrmion under $S$ is a 6-dimensional
manifold which we denote  by ${\cal M}_1$.
This is the manifold which is used as the moduli
space in the $B=1$ sector.
 It is diffeomorphic to
${\bf R}^3 \cross SO(3)$ and therefore its elements, which are called
Skyrmions, are   fully  specified
by their position and orientation.
On ${\cal M}_1$ the potential energy is constant and equal to the
Skyrmion's rest mass $M$, which can be calculated from the radial
function $f$. The result is $M=73 f_{\pi}/e$ \cite{ANW}.
Thus  the induced dynamics  is determined  by the kinetic
energy, which can be found by inserting the ansatz
\bea
\label{ans1}
A(t) U(\bx-\bX(t))A^{\d}(t)
\eea
into (\ref{KE}). After a lengthy calculation one finds the induced
Lagrangian \cite{ANW}
\bea
\label{Lsk1}
L_1 = -M + \Lambda\tr\dot{ A} \dot{ A}^{\d}   +
 {1\over 2}M \dot{\bX}^2,
\eea
where the dot denotes differentiation with respect to $t$.
The moment of inertia $\Lambda$ is also calculated in
 ref. \cite{ANW}  (where $\Lambda$ is denoted by  $\lambda$) and found
to be $53.3/e^3f_{\pi}$.
The  angular velocity $\bal$ is usually defined via
$A^{\dagger}\dot{ A }= -{i \over  2}\bal \cd \btau$.
In terms of the $SO(3)$ matrix $\cal A$ associated to  $A$
one can also write $
{\cal A}^{-1}\dot{\cal A} = \bal\cd \bs $
where $s_i$, $i=1,2,3$ are the generators of the $SO(3)$ Lie algebra
satisfying $[s_i,s_j] = \epsilon_{ijk}s_k$.
In terms of $\bal\>$,
$ \Lambda\tr\dot{ A} \dot{ A}^{\d} ={1\over 2}\Lambda \bal ^2$.
According to the equations of motion derived from
$L_1$      a Skyrmion moves uniformly in space and rotates  rigidly at
constant angular velocity.

So far the standard, approximate treatment of the $B=1$ sector.
How useful are the approximations made so far?
Clearly the uniformly moving but non-rotating solution
is a good approximation to the exact solution one obtains by
Lorentz-boosting a static Skyrmion, provided the speed is small
and Lorentz contraction can be neglected.
However, in this case the moduli space approximation is merely
a complicated way of deriving something trivial. The other,
rigidly rotating solution found in  the moduli space approximation
is almost certainly a poor approximation  to solutions of the
full field equations.
This can be seen in a number of ways.
One  argument,  frequently  discussed in the literature and
reviewed in ref.  \cite{HS}, is based on the observation that there
are centrifugal effects arising from the rotation which could pull
the Skyrmion apart.
We prefer to discuss the rotating Skyrmion from another point
of view which suggests that  an exact solution exists
 only if  $m_{\pi}\neq 0$ and that it  is not of the hedgehog form at all.
Our discussion is based on the following  simple observations.
Far away from the centre of the Skyrmion the field equations
(\ref{EL}) simplify to three uncoupled  free scalar wave
equations for the massless pion fields
\bea
\label{KG}
{\partial^2 \pi^a \over \partial t^2} -\Delta \pi^a = 0 \qquad a= 1,2,3.
\eea
Furthermore the  pion fields of  a   Skyrmion $\> AU_{H}(\bx)A^{\dagger}\>$
centred at  the origin behave asymptotically like
\bea
  \pi^a \sim
f_{\pi}\lambda {{\cal A}^{aj}
                        \hat{x}^j \over r^2}
   =  f_{\pi}\lambda {{\cal A}^{ij}e_a^i \hat{x}^j \over r^2}
\eea
where $\lbrace \be_a\rbrace$ is the standard basis of
${\bf R}^3$.
Defining
\bea
\label{dip}
\bp_a  = { 4\pi f_{\pi}\lambda } {\cal A}^{-1} \be_a
\qquad \mbox{(so that} \quad \pi^a \sim {\bp_a\cd\hat\bx \over 4\pi r^2}\>)
\eea
we see that for sufficiently large $r$ each of the pion fields
$\pi^a$ can be approximated by a static solution of the
scalar wave equation with  a suitable
(time-independent) dipole source term
(we will discuss the charge distribution for a scalar dipole systematically
 in sect. 6):
\bea
\label{dipeq}
{\partial^2 \phi \over \partial t^2} -\Delta \phi =
 -{\bp}\cd\nabla \delta^{(3)}(\bx).
\eea
This suggests that the asymptotic behaviour of each  pion field of a
rotating Skyrmion should be described by solutions of
(\ref{dipeq}) with rotating dipole moments ${\bp}_a$.
More precisely,
 for a dipole moment $\bp(t)$
rotating at constant angular velocity $\bal$, i.e.
$\dot\bp(t) = \bp(t)  \cross \bal $,
equation  (\ref{dipeq}) is solved by
\bea
\label{exact}
\lefteqn{
\phi(t,\bx)
= {1\over 4\pi r^2}[(\hat{\bx} \cd \bp_{\parallel}(t) +
(\cos \alpha r + \alpha r\sin\alpha r) \hat{\bx}\cd \bp_{\perp}(t)
}
\qquad \qquad \qquad
\nonumber \\
&&
+(\alpha r\cos \alpha r - \sin \alpha r) \hat{\bx}\cd
( \bp(t)  \cross \hat{ \bal})]
\eea
where $\alpha =|\bal|$ and $\bp_{\parallel}$ and $\bp_{\perp}$ are the
components of $\bp$ parallel and perpendicular to  $\bal$.
This should be compared with the asymptotic pion fields of the rigidly
rotating Skyrmion
\bea
\pi^a(t,\bx)  \sim { \bp_a(t)\cd\hat{\bx}  \over 4\pi r^2 }.
\eea
Clearly the rigidly rotating solution is only adequate if $r<1/\alpha$.
Furthermore we calculate, using standard formulae from radiation
theory, that the exact solution (\ref{exact}) will
radiate away its energy at a rate
\bea
\label{power}
P = {1\over 12 \pi} \alpha^4 |\bp|^2.
\eea
This suggests  that  a spinning Skyrmion should similarly lose
energy  through radiation and we can use the formula (\ref{power})
to estimate the time for which the approximation of uniform
rotation is  good.
To sum up, the dipole picture indicates  that,
for $m_{\pi} = 0$,  a rigidly  and
uniformly rotating Skyrme field can only be  a good
approximation to an exact solution of the Skyrme field
equations for times less than
\bea
t_{\mbox{\tiny life}} = {{1\over 2}\Lambda \alpha^2 \over P} =
{6 \pi \Lambda \over |\bp|^2 \alpha^2}
\eea
and for distances from the centre of the Skyrmion less
than
\bea
r_{\mbox{\tiny max}} = {1\over \alpha}.
\eea

Which value should we insert for the frequency $\alpha$ in order to
obtain numerical estimates for the above parameters?
Spin $1/2$ quantum states of a single Skyrmion are models for physical
baryons and spin $3/2$ states model the $\Delta$ - resonance.
We
fix the constants $f_{\pi}$ and $e$ as in ref. \cite{ANW} to
fit the nucleon and $\Delta$ masses
\bea
\label{paras}
f_{\pi} = 65\mbox{MeV} \qquad \mbox{and } \qquad e= 5.45,
\eea
so that $f_{\pi}$ is 30$\%$ smaller than its usual value of $93$MeV.
Thus the frequency at which a classical Skyrmion would have
to rotate to have the proton's or neutron's angular momentum is
\bea
\label{protfreq}
\alpha_N \approx {1\over 2\Lambda}\approx 1.5 f_{\pi} \approx 100\mbox{ MeV}.
\eea
We then find that
\bea
  t_{\mbox{\tiny life}} \approx {3 \over f_{\pi}}\approx 3\cross 10^{-23}s
\qquad \mbox{and} \qquad
r_{\mbox{\tiny max}}\approx {0.7 \over f_{\pi} } \approx 2 \>\mbox{fermi}.
\eea

A classical Skyrmion spinning
at the frequency $\alpha_N$
would therefore radiate away its rotational energy in a very short time
if $m_{\pi}= 0$, but it could be stabilised by including the
physical pion mass since
$ \alpha_N  < m_{\pi}$.
Note, however,  that in   a fully quantised
form  of the Skyrme model
the short classical lifetime   need not necessarily
be a cause for concern:
a  proton or neutron could only lower its rotational energy by emitting
a particle of spin $1/2$, but there is no such particle
in the model.

The value $r_{\mbox{\tiny max}}$ should be compared with
the root-mean-square
radius of  the Skyrmion's mass distribution, which is
$r_{\mbox{\tiny}}  = 0.7$ fermi \cite{HS}.
The form of the   pion fields   at  $r> 2$ fermi  does not
matter much in the calculation of the static nucleon properties
in the Skyrme model, but it is crucial for
 the long-range interaction of two Skyrmions.
In  studying the long-range forces between spinning Skyrmions, and
in particular in applying the Skyrme model to the nuclear two-body
problem, one  therefore needs to bear in mind that,
no matter whether $m_{\pi}=0$ or $m_{\pi}\neq 0$,
the long-range fields of a spinning Skyrmion
are not of the hedgehog form.

Even if we restrict ourselves to circumstances where  the rigidly
rotating hedgehog field  is an adequate approximation it does
not follow that a moving and spinning Skyrmion is well described by
(\ref{ans1}).
Rather, one should Lorentz-boost a stationary and rigidly rotating
hedgehog field to obtain a moving and spinning Skyrmion.
More precisely, suppose that the Skyrmion
is moving with velocity $\bu$ and has position $\bX$ at time
$t$. Its orientation at time $t$ is given by an
$SU(2)$ matrix $A$, and  the orientation then changes
 uniformly with respect to
the {\it rest frame } time, which we denote by $t'$.
We set  $t'=0$ at the space-time point with
laboratory coordinates
$(t,\bX)$; at a later   laboratory
time $t+ \delta t$, it is given by
\bea
t' = \gamma (\delta t - \bu\cd(\bx-\bX))
\eea
where  $\gamma = (1-\bu^2)^{-1/2}$. Thus the field for a moving and
spinning Skyrmion is
\bea
\label{ans2}
U(t +\delta t,\bx) = A(t')U_H(\bx')A^{\dagger}(t'),
\eea
where
\bea \bx_{\parallel}' = \gamma (\bx_{\parallel} -\bX_{\parallel} -\bu \delta t)
\quad \mbox{and}
 \quad
\bx_{\perp}' = \bx_{\perp} - \bX_{\perp} \nonumber .
\eea
Here   the  suffixes $\parallel$ and $\perp$ denote the
components parallel and perpendicular to the velocity vector $\bu$
and $A(t')$ is an  $SU(2)$-valued function defined by
\bea
\label{restan}
 A(0)=A \qquad \mbox{and} \qquad
-{i\over 2} \bal\cd\btau = A^{\dagger}(t'){d A (t')\over d t'}.
\eea
where $\bal$ is a constant vector, representing the angular velocity in
the Skyrmion's rest frame.
The Skyrme field (\ref{ans2}) takes into account Lorentz contraction
of a moving Skyrmion and the relativity of time: if the pion fields' rotation
is synchronised in the Skyrmion's rest frame then their rotation will be
retarded or advanced in the laboratory frame, depending on their distance
from the Skyrmion's centre projected onto $\bu$.

If the velocity $\bu$ is constant the field (\ref{ans2}) is a good
description of a spinning and moving Skyrmion for all laboratory times
$t+\delta t$.
Later in this paper we will consider Skyrmions which do not move uniformly
but whose acceleration is small. Then one needs   to perform different Lorentz
boosts
for different laboratory times and the formula (\ref{ans2})
is only an approximation, valid for small increments $\delta t$.
This is sufficient for us, since we will only require Skyrme fields and
their rate of change (with respect to laboratory time) at a given laboratory
time $t$.
  The position vector $\bX$ and the
orientation matrix $A$ strictly depend on $t$  but, for ease of notation,
we will not  write this dependence explicitly. Note also that
differentiation with respect to time should be carried out according to
\bea
\label{diff}
{\partial U \over \partial t}(t,\bx) = { \partial U(t +\delta t,\bx) \over
\partial \delta t}|_{\delta t=0}.
\eea
In our calculations will only take  into account the velocity dependence
of the field (\ref{ans2}) to leading order.
To this accuracy, and with the  limitations discussed
earlier, we will
make the approximation
\bea
\label{timeapp}
t' \approx \delta t - \bu\cd(\bx -\bX)
\eea
and  describe a single
moving and spinning Skyrmion by the following field:
\bea
\label{ans3}
U(t+\delta t,\bx) = A(t') U_H(\bxi) A^{\dagger}(t')
\eea
where
\bea
\bxi &=& \bx-\bX-\bu \delta t+
{1\over 2} (\bu\cd(\bx -\bX)) \bu \nonumber
\eea
and  it is implied that we use the  approximation
\bea
A(t')\approx A\left(\bone -{i\over 2}\bal\cd\btau(\delta t -\bu\cd(\bx-\bX))
\right).
\eea
When evaluating the Skyrme Lagrangian on fields of this form we
only retain terms which are at most quadratic in linear and angular
velocities.
Remarkably, the  evaluation of the Skyrme Lagrangian on the field
(\ref{ans3})
again yields again the Lagrangian $L_1$, but now  the angular velocity
$\bal$ should be interpreted as the angular velocity  in the rest frame
(\ref{restan}).

For later use we introduce the momenta
\bea
\label{spins}
\bJ = \Lambda \bal \quad \mbox{and} \quad \bI = \Lambda {\cal A} \bal
\eea
 which are the conserved quantities due to the invariance of $L_1$
under right-translations $A\mapsto AF$ and left-translations
$A \mapsto FA$ by a constant $SU(2)$ matrix $F$ respectively.
In the usual quantisation of a single Skyrmion $\bJ$ becomes the
spin operator and $\bI$ becomes the
 isospin operator.

\section{The Product Ansatz for Moving and Spinning Skyrmions}

In this section we outline our method for calculating the asymptotic form
of the Skyrme Lagrangian when evaluated on fields of the product form.
This method applies equally to the standard  product ansatz, which
employs fields of the form (\ref{ans1}) for the individual Skyrme fields,
and to our relativised product ansatz, which uses the fields (\ref{ans3}).
For definiteness, let us consider the latter.

Assume that, at time $t$,  one Skyrmion has  orientation
 $A\in SU(2)$,
angular velocity in its  rest frame $\bal$ , position $\bX$, and
velocity $\bu$ and the other Skyrmion has orientation   $B\in SU(2)$,
angular velocity in its rest frame $\bbet$,  position $\bY$ and  velocity
$\bv$. Then we write  the relativised product ansatz at time $t+\delta t$ close
 to $t$ for the field
describing this situation as
\bea
\label{proda}
U(t+\delta t,\bx) = U^{(1)}(t+\delta t,\bx)U^{(2)}(t+\delta t,\bx).
\eea
Here  $U^{(1)}$ is given by (\ref{ans3}) and $U^{(2)}$ analogously by
\bea
\label{ans4}
U^{(2)}(t+\delta t,\bx) = B(t'') U_H(\bet) B^{\dagger}(t'')
\eea
where
\bea
t'' &=& (1-\bv^2)^{-{1\over 2}}(\delta t - \bv\cd(\bx-\bY))\approx\delta t -
\bv \cd (\bx - \bY) \nonumber \\
\bet &=& \bx-\bY-\bv \delta t+
{1\over 2} (\bv\cd(\bx -\bY)) \bv \nonumber
\eea
and  the function $B(t'')$ is approximated by the following expression,
adequate for our purposes:
\bea
B(t'')&\approx& B\left( \bone - {i\over 2}\bbet\cd\btau ( \delta t -
\bv\cd(\bx -\bY))\right).
\eea

When inserting the field (\ref{proda}) into the Skyrme Lagrangian we first
 carry out all necessary differentiations  and then set $\delta t=0$.
For our calculations it will also be useful to introduce
\bea
\bR = \bX- \bY,  \; R = |\bR|
\eea
for the relative position.
The computation of the Lagrangian is simplified by the observation that
\bea
L_{\mu} = U^{\d}\partial_{\mu}U = U^{(2)\d}(\L_{\mu} - \R_{\mu} )U^{(2)}
\eea
where  $ \L_{\mu} = U^{(1)\d}\partial_{\mu}U^{(1)}$
 and  $ \R_{\mu} = U^{(2)}\partial_{\mu}U^{(2)\d}$.
When inserting this expression into (\ref{Lag}) the conjugation with
$U^{(2)}$ drops out. One still generates a large number of terms,
but they can be further classified according to their degree in
$\L_{\mu}$ or $\R_{\mu}$. The terms which involve only either $\L_{\mu}$
or $\R_{\mu}$
give the sum of the Lagrangians for a single free Skyrmion.
When expanding the remaining terms in the Lagrangian in powers
of $1/R$,
terms which are linear in $\L_{\mu}$ or
$\R_{\mu}$ give the leading contribution. Neglecting the terms which are
quadratic in both $\L_{\mu}$ and  $\R_{\mu}$  we get
\bea
 {\cal L} &=& -{f_{\pi}^2 \over 4 } \tr (\L_{\mu}L^{(1)\mu})
 + {1 \over 32 e^2}
\tr(\lbrack \L_{\mu},\L_{\nu}\rbrack\lbrack L^{(1)\mu},
L^{(1)\nu}\rbrack )
\nonumber \\
     &-&  {f_{\pi}^2\over 4 }\tr(\R_{\mu}R^{(2)\mu}) + {1 \over 32 e^2}
\tr(\lbrack \R_{\mu},\R_{\nu}\rbrack\lbrack R^{(2)\mu},
R^{(2)\nu}\rbrack )
\nonumber \\
     &+& {f_{\pi}^2\over 2 }\tr \left(\L_{\mu}\bar{R}^{(2)\mu} +
\Lbar_{\mu} R^{(2)\mu} - \L_{\mu}R^{(2)\mu}\right).
\eea
In terms of the mass $M$ and the moment of inertia $\Lambda$ of a single
hedgehog Skyrmion as defined earlier (\ref{Lsk1})
we can write the Lagrangian restricted to  fields of the product
form (\ref{proda})
\bea
L_2 = -2M + {1\over 2}M\bu^2 + {1\over 2}\Lambda \bal^2 + {1\over 2}M\bv^2
+ {1\over 2} \Lambda \bbet^2 + L_{\mbox{\tiny int}}
\eea
where
\bea
\label{Lint}
L_{\mbox{\tiny int}} = {f_{\pi}^2\over 2 }
\int d^3x\> \tr \left(\L_{\mu}\bar{R}^{(2)\mu} +
\Lbar_{\mu}R^{(2)\mu} - \L_{\mu}R^{(2)\mu}\right).
\eea
Our aim is to
calculate the leading terms in   an expansion of
 $L_{\mbox{\tiny int}}$ in powers of $1/R$.
Our calculations are based on the  idea of treating the field of
one soliton as constant in the vicinity of the other. In order
to evaluate the integral (\ref{Lint}) we divide the region of integration
into three parts
\bea
\label{divide}
{\bf R}^3 = B_{\rho}(\bX) \cup B_{\rho}(\bY) \cup V_{\rho}(\bX,\bY)
\eea
where $B_{\rho}(\bX)$ is the ball of radius $\rho$ around $\bX$ and
\bea
V_{\rho}(\bX,\bY) ={\bf R}^3 -( B_{\rho}(\bX) \cup B_{\rho}(\bY)).
\eea
With this division we can exploit the simple asymptotic form of the
hedgehog profile.
Equation (\ref{asymf}) implies that, for large $|\bxi|$,
\bea
\L_{\mu}(t,\bx) \sim  \l_{\mu}(t,\bx):=
i\lambda \partial_{\mu}
\left( A(t'){\hat \bxi\cd \btau \over
                         |\bxi|^2}A^{\d}(t')\right)_{|\delta t=0}
\eea
and similarly   for  large $|\bet|$
\bea
\R_{\mu}(t,\bx) \sim  \r_{\mu}(t,\bx):=
-i\lambda \partial_{\mu}
\left( B(t''){\hat \bet \cd \btau \over
                         |\bet|^2}B^{\d}(t'')\right)_{|\delta t=0}.
\eea
Since $\L_{\mu}$  and $\Lbar_{\mu}$  differ only by a triple product
of the   $\L_{\mu}$'s  (similarly for $\R_{\mu}$) we also have, in the
above limits,
\bea
\Lbar_{\mu} \sim  \l_{\mu} \quad \mbox{and} \quad
\Rbar_{\mu} \sim  \r_{\mu}.
\eea
As $R$ becomes large we will choose $\rho$ so large  that outside
$B_{\rho}(\bX)$  and $B_{\rho}(\bY)$ we can replace
$\L_{\mu}$  and $\Lbar_{\mu}$ by $\l_{\mu}$, as well as
$\R_{\mu}$
and  $\Rbar_{\mu}$ by $\r_{\mu}$.
But we also want to keep $\rho$ small
enough so that $\l_{\mu}$ and $\r_{\mu}$ may be considered constant over
$B_{\rho}(\bY)$ and $B_{\rho}(\bX)$ respectively. We achieve this by letting
$\rho \rightarrow \infty$ as $R \rightarrow \infty$
in such a way that
$\rho / R\rightarrow 0$.
We then find
\bea
\label{Vasy}
 L_{\mbox{\tiny int}}   &\sim&  {f_{\pi}^2\over 2 }\lbrack
\tr \left( \r_{\mu}(\bX) \int_{B_{\rho}(\bX)} d^3x \>\bar{L}^{(1)\mu} \right) +
\tr\left( \l_{\mu}(\bY) \int_{B_{\rho}(\bY)} d^3x \>\bar{R}^{(2)\mu}
\right)\nonumber\\
&+&
\int_{V_{\rho}(\bX,\bY)} \tr\left( \l_{\mu} r^{(2)\mu}\right) \rbrack.
\eea
This formula is the starting point in most of our calculations in the
following sections.

\section{The Two-Skyrmion Potential Revisited}
As a first step we consider  the static situation, where $\bu=\bv=0$ and
$A(t')=A$ and $B(t'')= B$ are constant
$SU(2)$ matrices.
Then $-L_{\mbox{\tiny int}}$ reduces to the
 two-Skyrmion
potential $V_2$:
\bea
\label{pot}
V_2 &=& {f_{\pi}^2\over 2 }
\int d^3x\> \tr \left(\L_i\Rbar_i +
\Lbar_i \R_i - \L_i \R_i\right)\nonumber \\
&\sim&
 {f_{\pi}^2\over 2 }\lbrack
\tr \left( \r_i(\bX) \int_{B_{\rho}(\bX)} d^3x \>(\Lbar_i- \l_i) \right) +
\tr\left( \l_i(\bY) \int_{B_{\rho}(\bY)} d^3x \>(\Rbar_i - \r_i)
\right)\nonumber\\
&+&
\int_{{\bf R}^3} d^3 x \tr\left( \l_i \r_i\right) \rbrack.
\eea
 The asymptotic form of $V_2$  was already given  by Skyrme
in ref. \cite{Skyrprod} and the potential energy of static
product fields for any value of $R$   was studied numerically  in
ref. \cite{prod}. We will re-derive Skyrme's result using a different method
which
we find mathematically more satisfactory. In the course of our calculation
we will also  derive a number of formulae which we will need  in  the
remainder of the paper.

 First we introduce  the $SO(3)$ matrix
\bea
\O^{ij} = {1\over 2} \tr (\tau^i A^{\dagger}B \tau^j B^{\dagger}A)
\eea
 specifying the relative orientation of the two Skyrmions.
 The main input of our calculation is the fact that
both $U^{(1)}$ and $U^{(2)}$ satisfy the Euler-Lagrange
equation for static fields:
\bea
\label{static}
\partial_i \Lbar_i = \partial_i \Rbar_i = 0.
\eea
Let us for the moment concentrate on
 $\Lbar_i$ and write it as
\bea
\Lbar_i = iA \bar{L}_{im}\tau_m A^{\d}.
\eea
Then, for each $m=1,2,3$, $\bar{L}_{im}$ are the components of  a smooth vector
field on ${\bf R}^3$
and (\ref{static}) implies that this vector field  is divergenceless.
Thus, for each $m$  we can find a  vector field with components  $Z_{km}$ such
that
\bea
\bar{L}_{im} = \epsilon_{ijk}\partial_j Z_{km}.
\eea
Explicitly we find
\bea
Z_{km}(\bx +\bX) = \left( \sin^2 f(r) -{1 \over  e^2  f^2_{\pi}  }
       {\sin^4 f(r) \over r^2} \right)
\hat{x}_k \hat{x}_m
+
\left( {1\over 2} r f' (r) + {1\over e^2 f^2_{\pi} } f' (r)
{\sin^2 f(r)\over r}\right)
 \epsilon_{kmn}\hat{x}_n.  \nonumber\\
\eea
Hence
\bea
\label{formula}
   \int_{B_{\rho}(\bX)} d^3x \,\Lbar_i(\bx)
&=&i A \tau_m A^{\d}
\int_{\partial B_{\rho}(0)}dS\, \epsilon_{ijk}Z_{km}(\bx +\bX)\hat{x}_j
\nonumber \\
&=&i A \tau_m A^{\d} \int_{\partial B_{\rho}(0)}dS \,\left( (\delta_{im}
-\hat{x}_i\hat{x}_m)
({1\over 2} rf' (r) + {1\over e^2 f^2_{\pi}} f' (r)
{\sin^2 f(r)\over r} )\right)
\nonumber \\
&=& i A \tau_i A^{\d}
{4 \pi \over 3}\left(\rho^3 f'(\rho)
+ {2\over e^2 f^2_{\pi}}\rho f' (\rho)
\sin^2 f(\rho)\right)
\nonumber\\
&\sim&  -{8\pi \lambda i \over 3} A\tau_i A^{\d}.
\eea
In the last line we have used that $\rho$ is so large that we can use the
asymptotic expression (\ref{asymf}) for $f$ and have omitted
non-leading terms.
Combining this with
\bea    \int_{B_{\rho}(\bX)} d^3x \,\l_i(\bx)
&=&-i\lambda A \tau_m A^{\d}
 \int_{B_{\rho}(\bX)} d^3x\>
\partial_i\partial_m{1\over |\bx-\bX|} \nonumber \\
&=& -i \lambda A\tau_m A^{\d} {1\over 3}\delta_{im}
 \int_{B_{\rho}(0)} d^3x\> \Delta {1\over |\bx|}\nonumber \\
&=& {4 \pi \lambda i \over 3} A \tau_i A^{\d}
\eea
we have
\bea
 \int_{B_{\rho}(\bX)} d^3x \>(\Lbar_i- \l_i) \sim -4\pi\lambda i A\tau_i
A^{\d}.
\eea
Since
\bea
\r_i(\bX) = -i \lambda {\delta_{in} -3\hat{R}_i\hat{R}_n \over R^3}
 B\tau_n B^{\d}
\eea
we therefore find
\bea
\tr \left( \r_i(\bX) \int_{B_{\rho}(\bX)} d^3x \>(\Lbar_i- \l_i) \right)
\sim -8 \pi \lambda^2 {\tr \O -3\hat{\bR} \cd \O \hat{\bR} \over R^3}.
\eea
An analogous calculation for the integral over
$B_{\rho}(\bY)$ yields also
\bea
\tr\left( \l_i(\bY) \int_{B_{\rho}(\bY)} d^3x \,(\Rbar_i - \r_i) \right) \sim
- 8 \pi \lambda^2  {tr \O -3\hat{\bR} \cd \O \hat{\bR} \over R^3}.
\eea
To calculate
the remaining integral we integrate by parts and use again $\Delta (|\bx|^{-1})
= -4\pi\delta^{(3)}(\bx) $:
\bea
 \int_{{\bf R}^3} d^3x\> \tr \left( \r_i\l_i\right)
&=&2 \lambda^2 \O^{jk}  \int_{{\bf R}^3} d^3x
 \>\partial_i\partial_j{1\over |\bx-\bX|} \>
 \partial_i\partial_k{1\over |\bx-\bY|}  \nonumber\\
&=& 8\pi  \lambda^2
  {tr \O -3\hat{\bR} \cd \O \hat{\bR} \over R^3}.
\eea
Combining terms and introducing the constant
\bea
\label{kappa}
\kappa= 2\pi \lambda^2 f_{\pi}^2
\eea
we obtain the result:
\bea
\label{result}
V_2 &\sim&
 -2\kappa
{\tr \O -3(\hat{\bR} \cd \O \hat{\bR}) \over R^3}=
2\kappa (\nabla\cd\O\nabla){1\over R}.
\eea
Here $\nabla$ is the gradient operator with respect to $\bR$. In the following
we will use the symbols $\partial_i$ and $\nabla$ for differentiation with
respect to either $\bx$ or $\bR$. Which of the two possibilities applies
will be clear from the context.
Note that one would get $-V_2$ from integrating the
asymptotic expression
$\tr\left(\l_i\r_i\right)$
over the whole of ${\bf R}^3$. The effect of replacing the
singular part of $\l_i$ and $\r_i$ by the true, smooth solutions
inside the balls is to add  $2 \cross
V_2$.

One can usefully write $V_2$ by expressing the rotation
matrix $\O$ in terms of its  axis {\boldmath $ n$ } and the angle of rotation
about
that axis $\psi$, i.e.
\bea
\O_{ij} = \cos \psi \delta_{ij} + (1-\cos \psi)n_in_j
+ \sin \psi \epsilon_{ijk} n_k
\eea
  and finds
\bea
\label{poti}
V_2 \sim  -2 \kappa (1-\cos \psi)
{1 -3 (\hat {\bR}\cd \bn)^2  \over R^3  }   .
\eea

Although     the details of
our calculations may seem quite complicated, the essential steps are simple.
After suitably dividing the region of integration we were able to express
the integral for (\ref{Lint}) involving $U(\bx)$ at all points in ${\bf R}^3$
in the simple form (\ref{result})
which requires only information about
the asymptotic nature of the fields. This remarkable result rests
essentially on the fact that the two $B=1$ Skyrmions that enter the product
ansatz individually satisfy the equations of motion, as expressed in
(\ref{static}).

There is another, heuristic way of deriving (\ref{result}).
It is based on the observation, explained in sect. 3, that, from afar, a
Skyrmion looks like
a triplet of orthogonal scalar dipoles. Thus it is plausible that the
long-range
static potential between two Skyrmions can be understood in terms of
 scalar dipole interactions.
Specifically one considers
two  triplets $\lbrace \bp_a\rbrace$ and
$\lbrace \bq_a\rbrace$, $a=1,2,3$ of scalar dipoles.
There is
a scalar dipole interaction
between $\bp_a$ and $\bq_b$ only if $a=b$,
given by
\bea
{1\over 4 \pi} (\bp_a\cd \nabla )(\bq_a \cd \nabla) {1\over R}.
\eea
To obtain the total potential
we   simply sum the interactions of the pairs.
We also need the  formula  for the
dipole moments associated to a Skyrmion
in general orientation (\ref{dip}).
Let ${\cal A}$
and ${\cal B}$ be the $SO(3)$ matrices
 associated to the $SU(2)$ matrices
$A$ and
$B$ as in (\ref{orth}).
Then we have from (\ref{dip})
\bea
\label{translation}
\bp_a  = { 4\pi f_{\pi}\lambda } {\cal A}^{-1}\be_a
\qquad \mbox{and} \qquad
\bq_a  = {4 \pi f_{\pi}\lambda} {\cal B}^{-1}\be_a
\eea
so that
$\bp_a = \O \bq_a$. Thus
the potential for the interaction of two triplets of scalar
dipoles is
\bea
\tilde V_2 = 2 \kappa (\nabla \cd {\cal O} \nabla ){1\over R}
\eea
which agrees with the asymptotic form of $V_2$.

\section{Skyrmions as Triplets of  Scalar Dipoles}

In this section  we   take the dipole model further: we calculate the
velocity-dependent forces between two triplets of scalar dipoles.

How do moving scalar dipoles interact?
To answer this question we need to know the coupling between a massless
scalar field and  a general source and  to write down  the charge
distribution for a  moving and spinning scalar dipole.
We fix the coupling of a scalar field  $\phi$ to a general charge distribution
$\rho$ by writing the field Lagrangian density as
\bea
{\cal L}_{\mbox{\tiny  field }} = {1\over 2} \partial_{\mu} \phi \partial
^{\mu}
\phi + \rho \phi.
\eea
For given $\rho$ this leads to the wave equation
\bea
\label{EM}
{\partial^2 \phi \over \partial t^2} -\Delta \phi = \rho
\eea
which is solved by
\bea
\label{scalar}
\phi(t,\bx) = {1\over 4\pi}\int d^3x' {\rho(t-|\bx-\bx'|,\bx')
                \over |\bx -\bx'|}.
\eea
Turning to the charge distribution $\rho$ we  notice that it
must be a Lorentz scalar to make equation (\ref{EM}) consistent.
As  a first  step we give  the charge distribution of a moving point particle.
In its rest frame is is characterised by a coupling strength $g$ and
its position $\bX$ and is given by
\bea
\label{pointrest}
g\delta^{(3)}(\bx- \bX).
\eea
To find its charge distribution in a general frame we write
down a charge distribution which is manifestly a Lorentz scalar
and which reduces to    (\ref{pointrest}) in the particle's rest frame.
For a particle with world line $X(\tau) = (T(\tau),\bX(\tau))$
and four-velocity $u=(u^0,\bu)=dX /d\tau$
the required  distribution is
\bea
\label{point}
\rho_{\mbox{\tiny point }} (x)=
g \int\,d\tau\, \sqrt{u_{\mu}u^{\mu}}\delta^{(4)}(x - X(\tau)).
\eea
Carrying out the integration over the world line parameter $\tau$ and
using
\bea
\left( {dT\over d\tau} \right)^{-1}\sqrt{u_{\mu}u^{\mu}} =  \sqrt{1-\bu^2}
\eea
we find
\bea
\rho_{\mbox{\tiny point }}(t,\bx)= g \sqrt{1-\bu^2}\delta^{(3)}(\bx -\bX(t))
\eea
where, in an abuse of notation, we have written $\bX(t)$ for $\bX(T^{-1}(t))$.

Our model for a scalar dipole involves a world line $X(\tau)$ and
the   dipole moment  $\bp(\tau)$ in   the dipole's rest frame.
In that frame, which is characterised by
$u=  dX/d\tau =(1,0)$, we can think of the dipole as consisting
 of two equal and opposite
scalar point charges with rest charges $\pm g$ , situated at
$\bX + {1\over 2}\bd$ and $\bX -{1\over 2}\bd$, see fig. 1.

\vspace{1 cm}

\centerline{
\begin{picture}(300,300)(0,0)
\put (50,200){\vector(2,1){100}}
\put (35,195){\circle{20}}
\put (30,175){$-g$}
\put (25,195){\line(1,0){20}}
\put (165,255){\circle{20}}
\put (160,235){$+g$}
\put (155,255){\line(1,0){20}}
\put (165,245){\line(0,1){20}}
\put (150,25){\vector(-1,4){50}}
\put (120,250){$\bd$}
\put (90,180){$\bX$}
\end{picture}     }

\centerline{{\bf Fig.} 1}

\vspace{2cm}

\centerline{Dipole charge distribution in the dipole's rest frame}

\vspace{ 1cm}

Still in the rest frame we impose that $|\bd|$ remains constant during
time evolution and  introduce the  angular velocity $\bal$
\bea
{d \bd \over d \tau} =   \bd\cross\bal.
\eea
Then  the charge distribution depicted in fig. 1 is described by
\bea
g\sqrt{1-\left({d \bd \over d \tau}\right)^2}\>\delta^{(3)}(\bx-(\bX + {1\over
2}\bd))
          -g\sqrt{1-\left({d \bd \over d \tau}\right)^2}\>\delta^{(3)}(\bx-(\bX
- {1\over 2}\bd)).
\eea
In the limit
where $|\bd|$ becomes small  and $g$ large in such a way that the dipole
moment
\bea \bp = g\bd
\eea
remains constant this simplifies to
\bea
\label{charge}
- \bp\cd\nabla \delta^{(3)}(\bx -\bX).
\eea
Again we write this charge distribution in a general frame  by
 giving a Lorentz scalar distribution which reduces to (\ref{charge})
in the rest frame $u = (1,0)$:
\bea
\label{dipdist}
\rho_{\mbox{\tiny dipole}}(x)= -\int\,d\tau\,\sqrt{u_{\mu}u^{\mu}}
P_{\mu}(\tau)\partial^{\mu}
\delta^{(4)}(x - X(\tau)).
\eea
Here the four-vector $P$ reduces to the dipole moment
$\bp$ in the dipole's rest frame, so in particular  $u_{\mu}P^{\mu}=0$.
This condition determines $P$ in terms of $\bp$:
\bea
P^0 &=& \gamma (\bu \cd \bp) \nonumber \\
\bP &=& \bp + {\gamma^2 \over  \gamma +1} (\bu \cd \bp)  \bu
\eea
where   $\gamma = (1-\bu^2)^{-1/2}$ as before.
Making these replacements in (\ref{dipdist}) and carrying out the
integration over the world line parameter $\tau$ we arrive
 at the expression
\bea
\rho_{\mbox{\tiny dipole}} (t,\bx)= -[
{d \over dt } (\bu \cd \bp)  + {1\over  \gamma} \bp \cd \nabla
+ \left( {\gamma \over \gamma +1} - {1\over \gamma}\right)
(\bu \cd \bp)(\bu\cd \nabla) ] \delta^3(\bx -\bX(t)).
\eea
Here $\bp = \bp(T^{-1}(t))$ should be  thought of as a function of time.
Assuming $d\bu/dt \approx 0$ and retaining only terms which are at most
quadratic in the velocities
$\bu$ and $\bal$ this reduces to the distribution
\bea
\rho_1 (t,\bx)= -[(\bu \cd \bp \cross \bal)
+ (1-{1\over 2} \bu^2)\bp\cd \nabla - {1\over 2}(\bu\cd \bp)
(\bu \cd \nabla)]  \delta^3(\bx -\bX(t)).
\eea

We can now calculate the potential due to $\rho_1$ according
to (\ref{scalar}). Since  $\rho_1$
varies slowly with time for slowly moving sources
we can make some approximations:
\bea
\label{taylor}
\phi_1(t,\bx) \approx {1\over 4\pi}\left(
\int d^3x'{\rho_1(t,\bx')\over |\bx-\bx'| }
   -{d\over dt}\int d^3x'\rho_1(t,\bx')
+{1\over 2}{d^2\over dt^2} \int d^3x'\rho_1(t,\bx')|\bx'-\bx|\right).
\nonumber\\
\eea
We should be careful to understand the range of validity of our
approximation: for constant $\bp$ but time-varying $\bX$ we can neglect
higher order terms in the above Taylor expansion if $\bu$ is small
relative to the speed of light and the acceleration and higher
derivatives are small. For fixed $\bX$ and time dependent $\bp$, however,
higher terms in the Taylor series are not necessarily small even if
$\bp$ rotates uniformly. In that case
the dipole field is exactly given by (\ref{exact}) and (\ref{taylor})
is
 an expansion in powers of
$|\bal| |\bx-\bX|$  of the exact solution.
Retaining only terms that contain at most two time derivatives
we obtain
\bea
\phi_1(t,\bx) \approx {1\over 4\pi}\left(
(\bu\cd\bal\cross\bp) - (1-{1\over 2}\bu^2)
\bp\cd\nabla
+{1\over 2}(\bu\cd \bp) (\bu \cd \nabla ))
{1 \over |\bx -\bX|}
-{1\over 2} {d^2\over dt^2}\bp\cd\nabla |\bx-\bX|\right).\nonumber \\
\eea

Now consider a second scalar dipole with world line
$Y(\sigma)=(S(\sigma),\bY(\sigma))$ and with
dipole moment $\bq$ in its rest frame. Denote its three-velocity
$d \bY / dt$ by $\bv$ and its angular velocity in its rest frame
by $\bbet$, i.e. $d \bq / d\sigma = \bq \cross \bbet$.
Its charge distribution, in the same approximation as for the first dipole,
is
\bea
\rho_2(t,\bx) =
 -[ (\bv\cd\bq \cross \bbet) +(1-{1\over2}\bv^2)(\bq\cd\nabla)  - {1\over
2}(\bv\cd \bq)
(\bv \cd \nabla)]\delta^3(\bx-\bY(t)).
\eea
The part of the Lagrangian which describes the interaction of the
second dipole with the field produced by the first is therefore

\vbox{
\bea
\label{dipole}
\tilde{L}_{\mbox{\tiny int}}
 &=& \int d^3x'\rho_2(t,\bx')\phi_1(t,\bx') \nonumber\\
&\approx&
{1\over 4\pi}[
(\nabla\cd\bp)\,(\bv\cd\bbet\cross\bq) -
(\bq \cd\nabla)\,(\bu\cd\bal\cross\bp)\nonumber\\
&-& (1-{1\over 2}\bu^2 -{1\over 2}\bv^2)
(\bp\cd\nabla)(\bq\cd\nabla) \nonumber\\
&+&{1\over 2}(\bu\cd \bp)(\bu\cd\nabla)(\bq\cd\nabla)
+{1\over 2}(\bv\cd \bq)(\bv\cd\nabla)(\bp\cd\nabla)
]{1\over R}\nonumber \\
&-&{1\over 8\pi} (\bq\cd\nabla){d^2\over dt^2}(\bp\cd\nabla) R.
\eea}

\noindent As in the previous section $\bR= \bX-\bY$ , $R= |\bR|$.
 In the last line
$\nabla$ differentiates with respect to $\bR$ and $d / dt$, when
operating on $R $,  acts  on $\bX$ only.
It is useful to rewrite the last term
\bea
\label{parts}
 (\bq\cd\nabla){d^2\over dt^2}(\bp\cd\nabla) R &=&
{d \over dt}\left((\bq\cd\nabla){d\over dt}(\bp\cd\nabla) R\right)
\nonumber\\
&&-(\dot{\bq}\cd\nabla)( \dot{\bp}\cd\nabla) R
-(\dot{\bq}\cd\nabla)( \bp\cd\nabla)( \bu\cd\nabla) R
\nonumber\\
&& +(\bq\cd\nabla) (\dot{\bp}\cd\nabla) (\bv\cd\nabla) R
+(\bq\cd\nabla) (\bp\cd\nabla )(\bu\cd\nabla )(\bv\cd\nabla) R.
\nonumber \\
&&
\eea
The time derivative outside the big brackets in the first
line acts on all the dynamical variables $\bp,\bq,\bX,\bY$
so
\bea
{d \over dt}\left( (\bq\cd\nabla){d\over dt}(\bp\cd\nabla) R\right)
\eea
is a total time derivative which we may discard since we are working
with a Lagrangian.

Finally we translate our results into ``Skyrme language",
using the notation introduced in sect. 5.
Discarding the total time derivative in       (\ref{parts})
and summing over three pairs of dipoles $\bp_a$ and $\bq_a$
we find, using the formulae  (\ref{translation}),

\vbox{
\bea
\label{dipoles}
\tilde{L}_{\mbox{\tiny int}}& =&  \kappa [
 2(\nabla\cd\O (\bv \cross \bbet))-
2(\bu \cross \bal  \cd \O \nabla)
\nonumber \\
&+&
(\bu^2 +\bv^2)
(\nabla\cd\O\nabla)
+(\bu\cd\nabla)(\bu\cd\O\nabla)
+(\bv\cd\nabla)(\nabla\cd\O\bv) ]{1\over R}\nonumber\\
&+& \kappa [
(\nabla\cross\bal\cd\O( \nabla \cross \bbet))\nonumber\\
&+&(\bv\cd \nabla) (\nabla \cross \bal \cd \O \nabla)
-(\bu\cd\nabla) (\nabla \cd \O (\nabla \cross \bbet))\nonumber\\
&-& (\nabla \cd \O \nabla)(\bu \cd \nabla)(\bv \cd \nabla)]R
\nonumber\\
&-&2 \kappa (\nabla\cd\O\nabla){1\over R}.
\eea  }

Note
 that the terms in this Lagrangian can be classified into three types:
spin-spin couplings which couple the angular velocities $\bal$ and $\bbet$
to each other with strength proportional to $1/R$, spin-orbit couplings
which couple $\bal$ and $\bbet$ to $\bu$ and $\bv$  with strength proportional
to $1/R^2$ and orbit-orbit couplings which are quadratic in $\bu$ and   $\bv$
and which are of order $1/R^3$.
Before we discuss our result further we derive it from the relativised
product ansatz.

\section{Two-Skyrmion Dynamics according to the Relativised Product Ansatz}

In this section we evaluate the interaction Lagrangian $L_{\mbox{\tiny int}}$
on  the time-dependent relativised product ansatz up to and including terms
which are quadratic in the velocities $\bu$,$\bv,\,\bal$ and $\bbet$.
For the sake of the impatient reader,
who may wish to proceed immediately to the comments at the end of this
section, we announce already at this point
that the asymptotic form of our final result agrees, up to a total  time
derivative, with
the interaction Lagrangian $\tilde L_{\mbox{\tiny int}}$ calculated
in the previous section.

We begin with the following  kinetic terms  of the interaction Lagrangian:
\bea
\label{Tint}
T_{\mbox{\tiny int}} = {f_{\pi}^2 \over 2} \int d^3x\,
\tr\left(\L_0\Rbar_0 + \Lbar_0\R_0 - \L_0\R_0\right).
\eea
For time dependent fields, $L_{\mbox{\tiny int}}$ actually contains further
terms with explicit time derivatives in the double commutator
$[\L_0,[\L_0,\L_i]]$ contained in $\Lbar_i$ (similarly for $\Rbar_i$)
but for computational purposes it is best initially to concentrate
 on $T_{\mbox{\tiny int}}$. Since each term in $T_{\mbox{\tiny int}}$
explicitly contains
two time derivatives we can omit the relativistic corrections
in the product ansatz (\ref{proda}) when calculating (\ref{Tint}) since
we only retain terms  quadratic in the velocities anyway.
Thus we replace $t'$ and  $t''$  by $\delta t$,
$\bxi$ by $\bx-\bX -\bu \delta t$ and
$\bet$ by $\bx -\bY -\bv \delta t$  in (\ref{proda}).
To carry out the integration in (\ref{Tint}) we employ
the same technique as for the static terms, so
we  divide ${\bf R}^3$ as in (\ref{divide}),
 extend the
integration for the asymptotic expressions of the fields from
$V_{\rho}(\bX,\bY)$ to the whole
of ${\bf R}^3$, subtract  the integrals of the singular parts over the
balls and add the integrals of the true fields over the balls.
This time we will omit most of the details, but it is instructive to
write down the integrals of the asymptotic expressions explicitly.
These now read
\bea
\l_0(t,\bx) = i\lambda A\left(\epsilon_{ijk}\alpha_i\tau_j\partial_k
{1\over |\bx-\bX|} + u_i\tau_j\partial_i\partial_j {1\over |\bx-\bX|}
 \right) A^{\d}
\eea
and
\bea
\r_0(t,\bx) =
-i\lambda B\left(\epsilon_{ijk}\beta_i\tau_j\partial_k
{1\over |\bx-\bY|} + v_i\tau_j\partial_i\partial_j {1\over |\bx-\bY|}
 \right) B^{\d}
\eea
so that

\vbox{
\bea
\label{integrals}
{f_{\pi}^2 \over 2}\int_{{\bf R}^3} d^3 x\, tr\left(\l_0 \r_0\right)
&=&f^2_{\pi} \lambda^2 \lbrack
\epsilon_{ijk}\epsilon_{lmn}\O_{jm}\alpha_i\beta_l
\int d^3x \>\partial_k {1\over |\bx-\bX|}\>\partial_n {1\over |\bx-\bY|}
\nonumber\\
&+&\epsilon_{ijk}\alpha_i v_l\O_{jm}
\int d^3x \>\partial_k {1\over |\bx-\bX|}\>
\partial_l\partial_m {1\over |\bx-\bY|}
\nonumber\\
&+&\epsilon_{lmn}\beta_l u_i\O_{jm}
\int d^3x \>\partial_i\partial_j {1\over |\bx-\bX|}\>
\partial_n {1\over |\bx-\bY|}
\nonumber\\
&+&u_iv_l\O_{jm}
\int d^3x \>\partial_i\partial_j {1\over |\bx-\bX|}\>
\partial_l\partial_m {1\over |\bx-\bY|}
\rbrack.
\eea }

\noindent We use the first of the above integrals to illustrate our method
of integration:
integrating by  parts, shifting the origin to $\bY$ we get
\bea
\int d^3x\> \partial_k{1\over |\bx-\bX|}\>\partial_n{1\over |\bx-\bY|}
&=&-\int d^3x \>{1\over |\bx|}\> \partial_n\partial_k{1\over |\bx-\bR|}
\nonumber \\
&=&-\int d^3x\>
{1\over |\bx|}
 {\partial\over \partial R_n}{\partial\over \partial R_k}
{1\over |\bx-\bR|} .
\eea
 Next we would like to take the differentiation outside the
integral. To justify this interchange of limits
we consider the above integrals over a large , but
finite ball $B_{K}(0)$
with
radius   $K> R$,
 centered at the origin.
Then
\bea
\int_{B_{K}(0)} d^3x
{1\over |\bx-\bR|}{1\over |\bx|} = 4\pi K -2\pi R .
\eea
For the finite region of integration, the exchange of limits
is allowed, so
\bea
\int_{B_{K}(0)} d^3x \>{1\over |\bx|}\>
 {\partial\over \partial R_n}{\partial\over \partial R_k}
{1\over |\bx-\bR|}& =&
{\partial\over \partial R_n}{\partial\over \partial R_k}(4\pi\kappa -2\pi R)
\nonumber\\
&=& -2\pi\partial_k\partial_n R .
\eea
We see that, as long as $K >R$, the integral on the left is
actually independent of $K$, we can take the limit
$K\rightarrow \infty$ and  recover the original integral.
Evaluating the other integrals in (\ref{integrals}) in  a similar
fashion we  obtain

\vbox{
\bea
\label{tensor}
{f_{\pi}^2 \over 2}\int_{{\bf R}^3} d^3x \>\tr\left(\l_0 \r_0\right)
&=&\kappa \lbrack
(\nabla\cross\bal\cd\O( \nabla \cross \bbet))R \nonumber\\
&+& (\bv\cd\nabla)(\nabla \cross \bal \cd \O \nabla) R \nonumber \\
&-& (\bu\cd\nabla)(\nabla \cd \O (\nabla \cross \bbet))R\nonumber\\
&-& (\nabla \cd \O \nabla)(\bu \cd \nabla)(\bv \cd \nabla)R \rbrack.
\eea  }

How is this result affected when we take into account the true
form of $\Lbar_0$ and $\Rbar_0$? As in the static case,
this amounts to adding
\bea
\label{sub1}
{f_{\pi} \over 2}\tr \left( \r_0 (\bX)\int_{B_{\rho}(\bX)}d^3x \>(\Lbar_0-\l_0)
\right)
\eea
and
\bea
\label{sub2}
{f_{\pi} \over 2}\tr \left( \l_0(\bY)\int_{B_{\rho}(\bY)}d^3x
 \>(\Rbar_0 -\r_0)\right).
\eea
The net effect of this is to change the signs of some but not all
of the terms generated when the differentiation in (\ref{tensor})
is carried out.
The terms in the integrand of  (\ref{sub1})
which are proportional to $\bal$ and the terms in the integrand of
(\ref{sub2}) which are
proportional to $\bbet$ lead to spin-spin terms  which fall off like $1/R^2$
 and  to
spin-orbit terms which fall off like $1/R^3$.
Such terms are non-leading and we neglect them.
Thus for our purposes $\Lbar_0 = - u_i \Lbar_i$ and
$\Rbar_0 = - v_i \Rbar_i$. Hence we  can again  exploit the fact that
the vector fields   $\bar{\bL}^{(1)}$ and
$\bar{\bR}^{(2)}$ are divergenceless, but we do not  give the details here.
The result
is

\vbox{
\bea
T_{\mbox{\tiny int}} &\sim&\kappa \lbrack
(\nabla\cross\bal\cd\O( \nabla \cross \bbet))R \nonumber \\
&+& (\bv\cd\nabla)(\nabla \cross \bal \cd \O \nabla) R
 - 2 (\nabla \cross \bal \cd \O\bv) {1\over R} \nonumber\\
&-& (\bu\cd\nabla) (\nabla \cd \O (\nabla \cross \bbet))R
+ 2(\bu \cd \O (\nabla \cross \bbet)){1\over R}\nonumber\\
&-& (\nabla \cd \O \nabla)(\bu \cd \nabla)(\bv \cd \nabla)R
+2(\bu\cd \O \nabla)(\bv \cd \nabla){1\over R}
+2(\nabla \cd \O \bv)(\bu \cd \nabla){1\over R}
\rbrack.                                     \nonumber\\
&&
\eea
  }

Next we evaluate the potential energy $V_2$ (\ref{pot}) on the
relativised product ansatz (\ref{proda}),
again setting $\delta t=0$ after  the calculation.
 The velocity dependence of that
field will lead to velocity-dependent modifications
of the potential calculated in sect. 5.
As a first step we restrict attention to constant matrix functions
$A(t')$ and $B(t'')$, i.e. vanishing angular velocities.
We then introduce
a ``rest frame  current" with components $\L_{\xi_{i}}$
via
\bea
 \L_{\xi_i} =  U^{(1)\d}{\partial \over \partial \xi_i}U^{(1)}
\eea
and find
\bea
\label{currs}
\L_i =\L_{\xi_i}  + {1\over 2} u_i (u_j \L_{\xi_j}).
\eea
Further we define the current with components
\bea
\Lbar_{\xi_i} = \L_{\xi_i} -{1\over 4 e^2 f_{\pi}^2}[\L_{\xi_j},[\L_{\xi_j},
\L_{\xi_i}]]
\eea
which satisfies
\bea
{\partial \over \partial \xi_i} \Lbar_{\xi_i} =0.
\eea
Carefully taking into account the  explicit time derivatives in the double
commutator \\ $[\L_0,[\L_0,\L_i]]$ contained in $\Lbar_i$
one checks that
\bea
\Lbar_i = \Lbar_{\xi_i} + {1\over 2}u_i (u_j \Lbar_{\xi_j}).
\eea
Defining
\bea
\R_{\eta_i} = U^{(2)}{\partial \over \partial \eta_i}U^{(2)\d}
\eea
and $\Rbar_{\eta_i}$ analogously to $\Lbar_{\xi_i}$ we also have
\bea
\Rbar_i= \Rbar_{\eta_i} +{1\over 2}v_i(v_j\Rbar_{\eta_j}).
\eea
Then, when integrating $\Lbar_i(t,\bx) $ over
$B_{\rho}(\bX)$,  we express
$\Lbar_i$ in terms of $\Lbar_{\xi_i}$ according to  (\ref{currs}) and change
integration  variables
via
$d^3 x = (1-{1\over 2}\bu^2)d^3 \xi$.
Thus, to order $\bu^2$,
\bea
\int_{B_{\rho}(\bX)} d^3x \,\Lbar_i (t,\bx)
&=&((1-{1\over 2}\bu^2)\delta_{im} +{1\over 2}u_i u_m)
\int_{B_{\rho}(0)} d^3\xi \,\Lbar_{\xi_m}(t,\bxi)\nonumber \\
&=&-((1-{1\over 2}\bu^2)\delta_{im} +{1\over 2}u_i u_m)
{8\pi \lambda i \over 3} A\tau_m A^{\d}
\eea
where we used (\ref{formula}) in the last line.
The integral of $\Rbar_i(t,\bx)$
over $B_{\rho}(\bY)$ can be dealt with  in an analogous fashion.
To calculate the remaining integrals in $V_2$
 we   need expressions for the asymptotic form of $\L_i$
and $\R_i$:
\bea
\label{compl}
\l_i(t,\bx) =
-i\lambda A\tau_m A^{\d}
[ \left( (1-{1\over 2}\bu^2) \partial_i\partial_m -{1\over 2} (\bu\cd\nabla)
u_m\partial_i \right) {1\over |\bx -\bX|} +{1\over 2}
(\bu\cd\nabla)(\bu\cd\nabla)
\partial_i\partial_m|\bx-\bX|] \nonumber\\
.\eea
and
\bea
\label{compr}
\r_i(t,\bx) = i\lambda B \tau_m B^{\d}
[\left( (1-{1\over 2}\bv^2)\partial_i \partial_m -{1\over 2}(\bv\cd\nabla)
v_m\partial_i \right) {1\over  |\bx -\bY|}
+{1\over 2}(\bv\cd\nabla)(\bv\cd\nabla)\partial_i\partial_m |\bx-\bY| ].
\nonumber \\
\eea
Using these formulae it is not difficult to reduce the integrations
of $\l_i$ , $\r_i$ and $\tr (\l_i \r_i)$
to integrals of the same form as those appearing in $T_{\mbox{\tiny int}}$
earlier
in his section.
The  resulting asymptotic expression for $V_2$ is
\bea
V_2\sim 2 \kappa
[(1-{1\over 2}\bu^2 -{1\over 2}\bv^2)(\nabla\cd \O \nabla)
+{1\over 2}(\bu\cd\nabla)(\bu\cd \O\nabla)
+{1\over 2}(\bv\cd\nabla)(\nabla\cd \O\bv) ]{1\over R}.
\eea

We still have to allow  for $A(t')$ and $B(t'')$ to vary with  $t'$
and $t''$
respectively, which leads to both being a function of  $\bx$.
This produces an additional term in the formula (\ref{compl}) for $\l_i$:
\bea
i\lambda A \partial_i\left(\bu\cd(\bx-\bX)(\bal\cross\nabla\cd \btau){1\over
|\bx -\bX| }\right)A^{\d}
\eea
and a similar extra term in the formula (\ref{compr}) for $\r_i$:
\bea
-i\lambda B\partial_i\left( \bv\cd(\bx-\bY)(\bbet\cross\nabla\cd\btau)
{1\over |\bx-\bY|}\right)B^{\d}.
\eea
Inserting these extra terms into (\ref{pot}) one thus
expects to obtain extra spin-orbit
terms of order $1/R^2$. Further spin-orbit terms are generated if one considers
the effect of the relativistic time transformation on $\Lbar_i$ and $\Rbar_i$,
but these
are of order $1/R^3$ and non-leading.
Miraculously, however, the  sum of the  extra  spin-orbit terms generated when
the
above corrections to $\l_i$ and $\r_i$ are taken into account
vanishes. Since all the integrals required to prove this result
have appeared earlier in this section we will not go through the details
here.
Physically  this result means that the relativity of time does not affect
the long-range interaction between two Skyrmions to the accuracy
considered here: we could have replaced $t'$ and $t''$ by $\delta t$
straight away.

Before we combine  $T_{\mbox{\tiny int}}$ and  $V_2$
into our final expression for $L_{\mbox{\tiny int}}$  we exploit
the fact that we can discard total derivatives in a Lagrangian.
Noting that, for any two vectors $\br$ and $\bs$
\bea
(\br\cd \dot\O\bs) =
(\bal \cross \br\cd\O \bs) +
(\br\cd\O(\bbet \cross  \bs))
\eea
and neglecting terms  proportional to $\ddot \bu$ (which are of order
$1/R^4$) we find that the  term
 \bea
[
( \bu\cd \O (\nabla \cross \bbet)) +
(\bu\cd\O \nabla)(\bv\cd \nabla)]{1\over R},
\eea
which occurs in $T_{\mbox{\tiny int}}$, can be replaced by
the  term
\bea
[
(\bal \cross \bu\cd \O \nabla) +
(\bu\cd \O \nabla)(\bu\cd \nabla) ] {1\over R}
\eea
because the two terms differ only
by the total derivative $
{d \over dt}( (\bu \cd \O \nabla)R^{-1})$.
Similarly we can replace
\bea
[
-(\nabla \cross \bal\cd \O\bv) +
(\nabla \cd \O \bv) (\bu\cd\nabla)]{1 \over R}
\eea
 by
\bea[
(\nabla\cd \O (\bv \cross \bbet)) + (\nabla\cd \O \bv)(\bv\cd\nabla)
]{1 \over R}
\eea
since the two terms differ by $
{d \over dt} \left( (\nabla \cd \O \bv)R^{-1}\right)$.

With these replacements we have  finally shown that the asymptotic form
of $L_{\mbox{\tiny int}}$ is
\bea
\label{RESULT}
L_{\mbox{\tiny int}}\sim \tilde L_{\mbox{\tiny int}}.
\eea

We already classified the terms in $ \tilde L_{\mbox{\tiny int}}$
at the end of the previous section, but the result (\ref{RESULT})
deserves some further comments.
First note that the Lagrangian
\bea
\tilde L_2 = -2M + {1\over 2}M\bu^2 +{1\over 2}M\bv^2
+{1\over 2}\Lambda\bal^2 +{1\over 2}\Lambda \bbet^2
+\tilde L_{\mbox{\tiny int}}
\eea
is not the asymptotic  Lagrangian which one would obtain in the usual moduli
space
approximation: we have evaluated the Skyrme Lagrangian $L$ (\ref{SKY}) on a
set of velocity-dependent fields which, at a given time, differ from fields
in the moduli space ${\cal M}^{12}$ by terms of order $\bu^2$ and $\bv^2$.
As a result the velocity-dependent terms in $\tilde L_2$ are not simply the
restriction of the kinetic energy $T$ to  the submanifold ${\cal M}^{12}$
of $\cal C$ but also include contributions from the potential energy evaluated
on the relativised,  and hence velocity-dependent product ansatz.

Our second comment concerns the validity of $\tilde L_2$ as an approximate
Lagrangian
for two-Skyrmion dynamics.
This is restricted in two ways. On the one hand it is well-known that
the product ansatz is only a good description of two well-separated
Skyrmions. To qualify ``well-separated" we note that the numerical work  in
ref. \cite{WW} indicates  that the product ansatz is already a good
approximation for Skyrmion
separations of $R\geq  1/2f_{\pi} \approx 1.5\>$ fermi.
On the other hand the allowed range of $R$ is restricted from above for
spinning
Skyrmions: for given individual angular velocities $\bal$ and $\bbet$ the field
of one Skyrmion at the position of the other is only well-described by the
rigidly rotating field implicit in the product ansatz if
\bea
\label{restriction}
|\bal| R <1 \quad \mbox{and} \quad |\bbet| R <1.
\eea
 This assumption was made explicitly in our di\-pole calculation.
When the individual Sky\-rmions are spinning at the nucleon frequency
$\alpha_N$ (\ref{protfreq}) the condition (\ref{restriction}) becomes
$R\leq 2\>$ fermi. In terms of  the standard classification of nuclear forces
according to their range \cite{EW}
this restriction means that the product ansatz, even in its relativised form,
can only be used to calculate the Skyrme model's predictions for
intermediate-range  nuclear forces. This  problem is routinely
ignored in  the literature; we  will come back to it in sect. 9.

Recall, finally, that effects due to the relativity of time turned out to be
irrelevant, at least to the accuracy considered here.
 This enabled us to  replace
the  rest frame  times $t'$ and $t''$ by the laboratory time $\delta t$.
Thus
we can, at a given time $t$, define the relative orientation $\O = {\cal
A}^{-1}{\cal B}$, which in turn  is essential for
the  notion of attractive channel fields, to be considered   in the next
section.

\section{Dynamics in the Most Attractive Channel}

The truncated Lagrangian system with Lagrangian $\tilde L_2$
is still rather complicated. To understand it better we
restrict attention to a particular set of fields
and truncate the Lagrangian system further.

It is clear from  (\ref{poti})
that the static two-Skyrmion potential is most attractive if one Skyrmion is
rotated by $180^{\circ}$ relative to the other about an axis orthogonal
to the line joining the two Skyrmions.
Such fields are said to be in the most attractive channel and they
can all be obtained by acting with the symmetry group $S$ (\ref{symmetry})
on a family of standard fields, parametrised by the separation parameter $R$.
 In terms
of a cartesian basis $\{\be_1,\be_2,\be_3\}$ of physical
space such a  standard field consists of
one standard hedgehog at ${1\over 2}R \be_3$ and another
hedgehog, rotated by $180^{\circ}$ about $\be_2$, at $-{1\over 2}R \be_3$:
\bea
\label{standard}
U_R(\bx)=U_{H}(\bx -{1\over 2} R \be_3) \tau_2 U_{H} (\bx +{1\over
2}R\be_3)\tau_2.
\eea
We can write the action of the group $E_3 \cross SO(3)_{\mbox{\tiny isospin}}
\in S$ in terms of  an $SU(2)$ matrix $G$ and its associated $SO(3)$ matrix
$\cal G$ for spatial rotations, a translation vector $\bS$, and  an $SU(2)$
matrix $C$ for isorotations.
Acting with $(G,\bS,C)$ on $U_R(\bx)$
we obtain
\bea
C G^{\d} U_H(\bx -\bS -{1\over 2} R {\cal G}\be_3)
G\tau_2 G^{\d}U_H(\bx -\bS +{1\over 2}R{\cal G}\be_3)G \tau_2 C^{\d}.
\eea
Comparing this expression  with our previous parametrisation
of product fields we find the following relationship between the
set $\{A,B,\bX,\bY\}$ and the set $\{G,\bS,C\}$:
\bea
A = CG^{\d} \qquad B= -iC\tau_2 G^{\d}
\nonumber \\
\bX = \bS +{1\over 2}R{\cal G}\be_3 \qquad \bY =\bS -{1\over 2}R{\cal G}\be_3.
\eea
Thus in particular $\hat \bR = {\cal G} \be_3$.
When $G,\bS$ and $C$  vary with time we also need notation for the
associated velocities.
For spatial rotations  we define the space-fixed (or right-invariant) angular
velocity $\bgam$ via
\bea
-{i\over 2}\bgam\cd\btau = \dot G G^{\d}
\eea
as well as the body-fixed (or left-invariant) angular velocity $\bom$
\bea
-{i\over 2}\bom\cd \btau = G^{\d} \dot G
\eea
which are related via ${\cal G} \bom =\bgam$.
Introducing  the left-invariant angular velocity $\bOm$ in isospace
\bea
-{i\over 2 }\bOm\cd\btau = C^{\d}\dot C
\eea
we find the following relationships:
\bea
\label{anvel}
\bal = {\cal G}(\bOm -\bom ) \qquad
\bbet= {\cal G}({\cal D}_2\bOm -\bom)
\eea
where ${\cal D}_i$ is the $SO(3)$ matrix for a rotation by $180^{\circ}$  about
the $i$-th axis.
For the linear velocities one checks that
\bea
\label{linvel}
\bu =  \dot \bS + {1\over 2}(\dot R \hat \bR + \bgam \cross \bR)
\qquad
\bv =  \dot \bS - {1\over 2}(\dot R \hat \bR + \bgam \cross \bR).
\eea
Using these relationships we shall now  express the restriction of $\tilde L_2$
to relativised product fields in the most attractive channel
in terms of $\bom,\bOm,\dot R,\dot \bS$ and  $R$.
This requires lengthy calculations which we will not describe in detail.
Note, however, that it is convenient to  use
\bea
\O_{ij} = -\delta_{ij} + 2 n_i n_j,
\eea
where $\bn = {\cal G}\be_2$, to simplify the terms in $\tilde L_2$.
One finds for example
\bea
(\bu \cd \nabla)(\bv \cd \nabla)(\nabla\cd\O\nabla)R
= -4{(\bu\cd\bn)(\bv\cd\bn) \over R^3}.
\eea
After simplifying $\tilde L_2$ in this fashion it is straightforward
to replace the velocities $\bal,\bbet,\bu$ and $\bv$ as in
(\ref{anvel}) and (\ref{linvel}).

Before we state the result we note some features which one can derive without
any calculations. Combining the inversion map $\Pi$ (\ref{inversion})
with rotations by $180^{\circ}$  in both space and isospace we obtain
$9$ isospace-space reflections
\bea
\Pi_{ai}: U(\bx) \mapsto \tau_a U^{\d}(\Pi_i \bx) \tau_a
\eea
where $\Pi_i$ is the reflection in the plane orthogonal  to $\be_i$.
Writing these maps more explicitly in terms of the pion fields
we have for example
\bea
\Pi_{13}: \bpi(\bx)\mapsto
(-\pi_1(x_1,x_2,-x_3),\pi_2(x_1,x_2,-x_3),\pi_3(x_1,x_2,-x_3)).
\eea
The maps $\Pi_{ai}$ are symmetries of the Skyrme Lagrangian and,
since they leave the set of product fields invariant,
they are also symmetries of the truncated Lagrangian system
with Lagrangian $\tilde L_2$.
In order to understand the action of  the maps $\Pi_{ai}$ on the
attractive channel fields it is sufficient to consider a
standard field $U_R(\bx)$ (\ref{standard}).
Such a  field is only strictly invariant under the action of
$\Pi_{13}$ but it also invariant under the action of
$\Pi_{11}$ and $\Pi_{22}$  modulo the commutator of two well-separated
hedgehog fields.
In our calculations of the asymptotic forces such commutators are
ignored. This can be seen from the fact that $\tilde L_2$ is symmetric under
the exchange of the individual Skyrmions position, orientation and linear
and angular velocities, although the product ansatz (\ref{proda}) is not
invariant under these operations. It was shown by Verbaarschot et al.
\cite{Verba} that the true attractive channel fields in standard
orientation and position, obtained by
minimising the static potential energy for a fixed value of $R$,
are in fact invariant under the
symmetries $\Pi_{13}$, $\Pi_{11}$ and $\Pi_{22}$.
This suggests that one should only trust the attractive
channel fields of the product form in the asymptotic limit,
where they  also have  those
symmetries.

The invariance (modulo commutator) of  $U_R(\bx)$
 under  $\Pi_{13}$, $\Pi_{11}$ and $\Pi_{22}$ implies
that, for each $R$, these maps act on the orbit of $U_R(\bx)$ under
 $E_3\cross SO(3)_{\mbox{\tiny isospin}}$.
Explicitly we find, for $(a,i)=(1,3),(1,1),(2,2)$:
\bea
\label{disc1}
\Pi_{ai}: \bS \mapsto \Pi_i \bS \quad
G \mapsto \tau_i G\tau_i \quad
C \mapsto \tau_a G\tau_a .
\eea
For the angular velocities this implies
\bea
\label{disc2}
\Pi_{ai}: \bom \mapsto {\cal D}_i \bom \quad \bOm \mapsto {\cal D}_a \bOm.
\eea
The restriction of $\tilde L_2$ to the orbit of the one-parameter family of
fields $\{ U_R(\bx)\}$
under $E_3 \cross SO(3)_{\mbox{\tiny isospin}}$, which we denote
by $L_{\mbox{\tiny att}}$, must be invariant
under (\ref{disc1}) and (\ref{disc2}) and also under the left action
of  $E_3\cross SO(3)_{\mbox{\tiny isospin}}$ on itself.
This implies that the most general form of $L_{\mbox{\tiny att}}$ in the
centre-of-mass frame ($\dot \bS =0$)
is
\bea
\label{at}
 \mu {\dot R}^2 + a^2 \omega_1^2 + d^2 \omega_2^2
+ c^2 \omega_3^2  + A^2\Omega_1^2
+B^2 \Omega_2^2 + C^2\Omega_3^2 +2e \omega_2\Omega_2
+{4\kappa \over R^3}
\eea
where $a,c,d,e,A,B,C$ and $\mu$ are functions of $R$ only.
For later use it is best to ``complete the square" and to rewrite this
expression slightly:
\bea
\label{att}
 \mu {\dot R}^2 + a^2 \omega_1^2 + b^2 \omega_2^2
+ c^2 \omega_3^2  + A^2\Omega_1^2
+B^2 (\Omega_2 + w  \omega_2)^2 + C^2\Omega_3^2
+{4\kappa \over R^3}
\eea
where $b^2 = d^2 -e^2/B^2$ and $w=e/B^2$.

Carrying out the explicit calculation of  $L_{\mbox{\tiny att}}$
we find
\bea
\label{Latt}
L_{\mbox{\tiny att}}&=& (M-{2\kappa \over R^3}) {\dot \bS}^2
-{6\kappa \over R^3}(\dot \bS \cd\hat \bR)^2
+({M\over 4} -{2\kappa \over R^3}) {\dot R}^2 \nonumber \\
&+&({ M \over 4}R^2 + \Lambda -{15\kappa \over  2R})\omega_1^2
+({M\over 4} R^2 +\Lambda -{7 \kappa \over 2R})\omega_2^2  + \Lambda\omega_3^2
\nonumber \\
&+&(\Lambda +{\kappa \over R})\Omega_1^2 +(\Lambda -{\kappa \over R})\Omega_2^2
+ \Lambda \Omega_3^2
- 2(\Lambda -{2\kappa \over R})\omega_2\Omega_2 +{4\kappa \over R^3}.
\eea
Note that this Lagrangian is not Galilei invariant.
Although the momentum $\bP_S$ conjugate to $\dot \bS$ is conserved
 one finds that in general $\ddot \bS \neq 0$.
We can, however, set $\bP_S =0$ and it then follows that $\dot \bS =0$.
 $L_{\mbox{\tiny att}}$  then takes the form (\ref{at}) and we read off the
coefficient functions $\mu$,
$a^2,c^2,A^2,B^2$ and $C^2$ directly; after bringing
it into the form (\ref{att}) we also find
\bea
b^2(R) ={M\over 4} R^2 -{ \kappa \over 2R} \quad \mbox{and}\quad
w(R)=-(1-{\kappa \over \Lambda R})
\eea
with corrections  of order $1/R^2$.

We remark that the kinetic part of $L_{\mbox{\tiny att}}$
can formally be interpreted in terms of a Riemannian structure on the
space of attractive channel fields. Such  a geometric interpretation
plays an important role in the  moduli space approximation to the dynamics
of Bogomol'nyi-type
solitons. In that context  the geometric viewpoint is emphasised
because the kinetic energy of the field Lagrangian can itself
be thought of in terms of a Riemannian structure on the configuration
space $\cal C$. It is  then consistent to interpret the kinetic energy
of the truncated Lagrangian in terms of an induced Riemannian structure.
Recall, however, that in our case  the kinetic part of
the Lagrangian $\tilde L_2$  is not just  the restriction of the
kinetic energy (\ref{KE}) to some submanifold of the configuration space $\cal
C$.
Thus the geometric viewpoint, though possibly interesting, seems less well
founded
in our case and we will not emphasise it in the following.

The asymptotic form of the functions $A^2,B^2,C^2$
\bea
\label{inertia}
A^2\sim B^2\sim C^2 \sim \Lambda
\eea
can easily be interpreted in terms of the Skyrmions' moments of
inertia. With our normalisation of the kinetic energy the formula
(\ref{inertia}) implies that the  asymptotic moment of inertia  in isospace is
$2 \Lambda$ about any axis, which is what one expects for the simultaneous
isorotation of two Skyrmions with moment of inertia $\Lambda$ each.
Similarly the asymptotic behaviour
\bea
\label{inert}
a^2 \sim b^2 \sim  {M \over 4} R^2  + \Lambda, \qquad  c^2 \sim \Lambda
\eea
can be understood in terms of the moments of inertia for the rigid
rotation of two Skyrmions  a distance $R$ apart:
$2\Lambda$ is the moment of inertia about the line joining the Skyrmions
and, by Steiner's theorem, $(2M)({R\over 2})^2 + 2\Lambda $  is the moment of
inertia about
either of the two perpendicular directions through the midpoint of the
line joining the Skyrmions and at right angles to it.

The corrections to the asymptotic expressions (\ref{inertia}) and (\ref{inert})
reflect the velocity-dependent forces between moving and spinning Skyrmions.
To study their effects on two-Skyrmion dynamics we will now briefly investigate
 the  equations of motion corresponding to
$L_{\mbox{\tiny att}}$, using the Hamiltonian formalism.
Thus we introduce body-fixed angular momenta  in space and isospace
\bea
\bm = {\partial L_{\mbox{\tiny att}} \over \partial \bom}
\quad  \bM = {\partial L_{\mbox{\tiny att}} \over \partial \bOm}
\eea
which are  conjugate to $\bom$ and $\bOm$ respectively; in components
\bea
m_1=2 a^2\omega_1, \qquad m_2=2(b^2 +B^2 w^2)\omega_2 +2B^2\Omega_2, \qquad
m_3=2 c^2\omega_3 \nonumber \\
M_1 = 2 A^2 \Omega_1, \qquad M_2 = 2B^2 (\Omega_2 +w\omega_2), \qquad
M_3=2C^2\Omega_3.
\eea
Finally defining $P_R = \partial L_{\mbox{\tiny att}} / \partial \dot R$
we arrive at the Hamiltonian in the centre-of-mass frame
\bea
H_{\mbox{\tiny att}} ={P_R^2 \over 4\mu}
+{m_1^2\over 4 a^2} + {(m_2 -w M_2)^2 \over 4 b^2} +{m_3^2 \over  4c^2}
+ {M_1^2 \over 4 A^2} + {M_2^2 \over 4 B^2} + {M_3^2 \over 4 C^2}
-{4\kappa \over R^3}.
\eea
The conserved quanitites are the Hamiltonian itself, the squared
spatial angular momentum
\bea
\bm^2 =m_1^2 + m_2^2 + m_3^2
\eea
and the squared isospin
\bea
\bM^2 =M_1^2 + M_2^2 + M_3^2
\eea
as well as the space-fixed angular momenta in space and isospace which
will, however, not concern us here.
Using the basic Poisson brackets
\bea
\{m_i,m_j\} = -\epsilon_{ijk}m_k ,\qquad
\{M_i,M_j\}= -\epsilon_{ijk}M_k, \qquad
\{m_i,M_j\} = 0
\eea
it is not difficult to write down the Hamiltonian equations of motion
$\dot F= \{H,F\}$ for
any function $F$ of the coordinates and momenta.
Exploiting the conservation laws one can  also study certain trajectories
qualitatively without integrating the equations of motion.

One can, for example, satisfy the equations of motion
by setting $\dot \bS =0$, $\bm = (m_1^0,0,0)$ and $\bM = (M_1^0,0,0)$,
where $m_1^0$ and $M_1^0$ are two
positive  constants, and then determine $R(t)$ from the energy conservation
law
\bea
E = ({M\over 4} - {2\kappa \over R^3}){\dot R}^2  +
{(m_1^0)^2 \over M R^2 +4\Lambda - 30 \kappa /R} + {(M_1^0)^2 \over 4\Lambda
+ 4\kappa /R} -{4 \kappa \over R^3}.
\eea
Using the numerical values for $M$ and $\Lambda$ given in
sect. 3  we can rewrite this equation in terms of
 the dimensionless quantity $\tilde R = f_{\pi}R$, and simplify it, assuming
$\tilde R >1$:
\bea
\label{bound}
{4E\over M}-{(M_1^0)^2 \over 4} = {\dot R}^2 + {1\over 45}\left({(m_1^0)^2
\over {\tilde R}^2}
-{(M_1^0)^2 \over  {\tilde R}}- {1.8 \over {\tilde R}^3}\right).
\eea
One  understands the time evolution of $R(t)$ qualitatively by looking
at the ``effective potential" in round brackets.
Remarkably this potential includes an isospin-dependent attractive Coulomb
potential
which could lead to a bound motion. For such a motion one finds that
$\tilde R \geq 2.3 / M_1^0$. However, the angular velocity
$\Omega_1 =M_1^0 /2 \Lambda$ leads, via (\ref{anvel}), to individual angular
velocities
$\alpha_1$ and $\beta_1$   which are also of the order $M_1^0 /2 \Lambda$ (one
checks
that $\omega_1$ is negligible) and which, by (\ref{restriction}), restrict the
validity of $L_{\mbox{\tiny
att}}$ to $\tilde R < 0.7/M_1^0$.
Thus the bound motion predicted by equation (\ref{bound}) is always outside
the range of validity of $L_{\mbox{\tiny att}}$.

Further studies of trajectories indicate that the result of our example
calculation is generic:
the velocity-dependent forces encoded in coefficient functions of
$L_{\mbox{\tiny att}}$ give only small corrections to the equations of
motion obtained from the following, simpler Lagrangian  in which the
velocity-dependent
forces are ignored.
\bea
L_{\mbox{\tiny stat}} &=& 2M\dot \bS^2 + {M\over 4} {\dot R}^2
+({ M \over 4}R^2 + \Lambda)\omega_1^2
+({M\over 4} R^2 +\Lambda)\omega_2^2  + \Lambda\omega_3^2 \nonumber \\
&+&\Lambda (\Omega_1^2 + \Omega_2^2
+  \Omega_3^2)
- 2 \Lambda \omega_2\Omega_2 +{4\kappa \over R^3}.
\eea
Trajectories calculated with $L_{\mbox{\tiny att}}$
are only qualitatively different from those calculated with
$L_{\mbox{\tiny stat}}$ if the momenta $\bm$ and $\bM$ are taken to be so large
that
$L_{\mbox{\tiny att}}$ is no longer a good approximation for two-Skyrmion
dynamics.

The calculations in this section and the previous one were based on moduli
space ideas, with some relativistic corrections.
The dipole picture suggests that this approach is fundamentally  limited
by the condition (\ref{restriction}) and consequently only capable of
calculating small velocity-dependent effects. However, the dipole model also
provides an
alternative  method
for studying the long-range forces between spinning Skyrmions
which  is in principle valid for any value of the Skyrmions' angular
velocities and any (sufficiently large) value of $R$.
This point of view is elaborated in the following section.

\section{Long-range Forces between Spinning Skyrmions and  Nuclear Forces }

The semiclassical angular speed $\alpha_N$ (\ref{protfreq}) at which a
Skyrmion has to rotate to have a nucleon's spin is so large  that the product
ansatz (even in the relativised form) is not a reliable basis for calculating
the Skyrme model's
prediction for the long-range nuclear force.
We will therefore give a heuristic derivation of
the long-range force between stationary but rapidly spinning Skyrmions based
on the dipole model.

Consider two triplets of well-separated and rapidly spinning
scalar dipoles.
The dipole moments
orthogonal  to the  direction of the angular velocity produce
a rapidly oscillating field  but the dipole moment in  the direction of the
angular velocity produces the simple  field of a static
dipole, cf. (\ref{dip}).
When two spinning triplets of dipoles interact the oscillating fields
produce no net force but the projections of the dipole moments
onto the direction of the angular velocities interact
like static dipoles.
According to the arguments of sect. 5 and  the formulae
(\ref{translation}) the potential energy for the interaction
 is therefore
proportional to
\bea
\label{spinnn}
\sum_{a=1}^3 (\bal\cd\bp_a)(\bbet\cd\bq_a)(\bal\cd\nabla)
(\bbet\cd\nabla){ 1\over 4\pi R}
= 2 \kappa
(\bal\cd{\cal O}\bbet)(\bal\cd\nabla)(\bbet\cd\nabla)
{1\over R}.
\eea
Recalling from (\ref{spins}) that the  spins for the individual Skyrmions
are given by
\bea
 \bJ^{(1)} = \Lambda \bal \quad \mbox{and } \bJ^{(2)}=\Lambda \bbet
\eea
and the isospins
by
\bea
\bI^{(1)}  = {\cal A}\bJ^{(1)} \quad \mbox{and} \quad
\bI^{(2)}  = {\cal B} \bJ^{(2)}.
\eea
we can write the potential energy for  the long-range
interaction between spinning triplets of dipoles as
\bea
 C\cross 2 \kappa
(\bI^{(1)}\cd \bI^{(2)})
(\bJ^{(1)}\cd\nabla)(\bJ^{(2)}\cd \nabla)
{1\over R}.
\eea
where $C$ is an unknown constant of proportionality.
Remarkably this potential has exactly the same isospin and spin
dependence  as the zero-mass version of the one-pion
exchange potential (OPEP) familiar from nuclear physics \cite{EW}.
In the usual treatment  of the nuclear two-body problem  in  the Skyrme model
one projects the static two-Skyrmion potential (\ref{result}) onto
tensor products of quantum states of free Skyrmions (which model nucleons)
\cite{prod}.
One then also reproduces the OPEP, which is  regarded as one of
the successes of the Skyrme model. Here we have offered an
alternative way of extracting the  OPEP from the Skyrme model which
takes into account the Skyrmions' spin and isospin already at the classical
level. One still needs to determine the constant $C$, however. This requires
more thought.
In particular   one  needs to
calculate the  dipole moments  of a
rapidly spinning Skyrmion.
Centrifugal effects could change them significantly relative to their
static values.

\section{Conclusion}

In this paper we have shown that a moduli space inspired technique
(the relativised product ansatz) and a point-particle model (the dipole
picture) lead to the same 12-dimensional Lagrangian system for asymptotic
two-Skyrmion
dynamics.
The interpretation of the agreement between the two approaches is
that well-separated Skyrmions interact like two point-particles
with suitable scalar dipole moments (adjusted to fit the asymptotic field of
a single Skyrmion) not only statically but also
dynamically.

Relativistic effects such as Lorentz contraction were important in our
calculation, and to incorporate them we had to modify the usual
moduli space prescription. In fact, on the basis of our calculations
one might be surprised
that in soliton theories of Bogomol'nyi type the moduli space
approximation without relativistic
corrections satisfactorily reproduces  the results of numerical calculations in
the full
field theory and  agrees
  asymptotically with  point-particle approximations.
The reason for this, however, is a special property  of soliton theories of
Bogomol'nyi type:
in such theories all fields in the moduli space are at
the minimum of the potential, and therefore Lorentz contraction leads
to corrections of order (velocity$)^4$.  By contrast,  in theories where
there are static inter-soliton forces, Lorentz contraction leads to
corrections of order (velocity$)^2$, i.e. of the same order as the kinetic
energy. In such theories  it is therefore not consistent to include the kinetic
energy
in a truncated Lagrangian but to ignore relativistic effects.

While we could modify the   moduli space approximation for two-Skyrmion
dynamics, as proposed
in \cite{M1},  to take into account Lorentz-contraction
of the individual Skyrmions,  we have argued that it has to be abandoned
altogether when studying the long-range forces between spinning Skyrmions.
In its stead
the dipole picture should be used to calculate those forces.
Moduli space techniques  are only reliable when the Skyrmions' separation
and their angular speeds satisfy the inequality (\ref{restriction})
 or when the  Skyrmions are so close
together that their interaction forces them to move collectively rather
than individually. However, for near-coincident or coincident Skyrmions
the product ansatz is no longer adequate and
the more sophisticated methods of ref.
\cite{AM} will be required. The Lagrangian $\tilde L_2$ would be useful
in this context because it  could
 interpolate between the moduli space and the point-particle approximation.

The immediate  question that arises from our work
 concerns the precise nature of spinning
Skyrmions. Here we have argued that the asymptotic field is that
of a spinning triplet of  dipoles,
 but it would be interesting to study spinning
Skyrmions numerically to check our argument and to calculate the field
near the Skyrmion's centre.
Such calculations should be carried out with massive pion fields, which
is necessary for the existence of a stable rotating solution of the Skyrme
equations. The results would flesh out or  refute the argument of sect. 9,
that it is the  forces between {\it spinning} Skyrmions which
one needs to understand in order to calculate nuclear
forces from the Skyrme model.

\vspace{1cm}

\vbox{

\noindent{\bf Acknowledgements} \\
This paper developed partly from work done while I was a research student at
DAMTP, Cambridge. I am grateful for discussions with Nick Manton and
Trevor Samols during that time. \\
I acknowledge an SERC research assistantship.
}

\end{document}